\documentclass[11pt]{article}
\usepackage{amsmath}
\usepackage{amssymb}
\usepackage{latexsym}
\usepackage{epsfig}

\newcommand{\be}{\begin{equation}}
\newcommand{\ee}{\end{equation}}

\newcommand{\Ci}{C}

\newcommand{\sectiono}[1]{\section{#1}\setcounter{equation}{0}}

\newcommand{\p}{\partial}

\begin{document}

{}~ \hfill\vbox{\hbox{hep-th/0502161}\hbox{MIT-CTP-3587} }\break

\vskip 3.0cm

\centerline{\Large \bf Dilaton  Deformations
  in Closed String Field Theory
}

\vspace*{10.0ex}

\centerline{\large Haitang Yang
and Barton Zwiebach}

\vspace*{7.0ex}

\vspace*{4.0ex}

\centerline{\large \it  Center for Theoretical Physics}

\centerline{\large \it
Massachusetts Institute of Technology}

\centerline{\large \it Cambridge,
MA 02139, USA}
\vspace*{1.0ex}

\centerline{hyanga@mit.edu, zwiebach@lns.mit.edu}

\vspace*{10.0ex}

\centerline{\bf Abstract}
\bigskip
\smallskip

The dilaton theorem implies that the contribution to the dilaton
potential from cubic interactions of all levels must be cancelled
by the elementary quartic self-coupling of dilatons.  We use this
expectation to test the quartic structure of closed string field
theory and to study the rules for level expansion.  We explain how
to use the results of Moeller to compute quartic interactions of
states that, just like the dilaton, are neither primary nor have a
simple ghost dependence. Our analysis of cancellations is made
richer by discussing simultaneous dilaton and marginal
deformations. We find evidence for two facts: as the level is
increased quartic interactions become suppressed and closed string
field theory may be able to describe arbitrarily large dilaton
deformations.

\vfill \eject

\baselineskip=16pt

\vspace*{10.0ex}

\tableofcontents

\sectiono{Introduction and Summary}\label{sect1}

Level-expansion computations in open string field theory
have been a useful tool in the study of open string tachyon
condensation~\cite{Taylor:2003gn}. The early attempts to
compute the potential for the (bulk) closed string tachyon
of bosonic strings~\cite{Belopolsky:1994sk,Belopolsky:1994bj}
were done before level expansion was
understood and the results were inconclusive.
Clearer results were obtained recently
in the computation of potentials for the twisted
tachyons~\cite{Okawa:2004rh,Bergman:2004st} that live on orbifold
cones. A workable level expansion scheme requires that
finite number of couplings be considered at each computation
stage. Since closed string field theory is nonpolynomial
it is not obvious that level expansion works. If
a class of closed-string computations
can be done in level expansion, it is then necessary
  to compute higher-order couplings efficiently.
The results of Moeller~\cite{Moeller:2004yy} make this possible
for the case of four-point couplings. Moeller has provided
the Riemann-surface data necessary to compute arbitrary
couplings of four string fields: a concrete description of the subspace
$\mathcal{V}_{0,4}\subset\mathcal{M}_{0,4}$ of the moduli space of
four-punctured spheres and the local coordinates around the
four punctures for every punctured sphere in $\mathcal{V}_{0,4}$.

In a previous paper~\cite{Yang:2005iu} we considered marginal
deformations in closed string field theory.  The marginal
parameter, called $a$, was that associated with the
dimension-zero primary operator
$c\bar c\p X \bar\p X$.  When the coordinate
$X$ lives on a circle the operator induces a change
of radius. The operator is marginal even when the coordinate
$X$ is noncompact, but adding it to
the action does not change the correlators of the conformal theory.
We used this marginal operator to test the quartic structure
of closed string field theory and the feasibility of level expansion.
We checked the vanishing of
the effective potential for $a$.   In the level expansion  the quartic
terms  generated by the cubic interactions (to all levels)
must be cancelled by the elementary quartic interaction
of four marginal operators. We confirmed this prediction, thus
giving evidence that the sign, normalization, and region of
integration $\mathcal{V}_{0,4}$ for the quartic vertex are all correct.
This was the first calculation
of an elementary quartic amplitude for which there was an
expectation that could be checked.
We also extended the calculation to the case of the four marginal
operators associated with two space coordinates.

In this paper we consider a  nontrivial extension of the above
results.  We study the potential
for  the zero-momentum dilaton, the field $d$ associated with
the operator ${1\over 2} (c\p^2 c - \bar c \bar\p^2 \bar c)$.
Complications arise because this operator is not  marginal:
it has dimension zero
but it is not primary since it fails to be annihilated by $L_1$ and
by $\bar L_1$.
The dilaton theorem~\cite{dilatontheorem}
   states that a shift in the expectation
value of $d$ corresponds to a change in the string coupling constant.
Around the flat spacetime background there is no potential for the dilaton,
so it behaves like a marginal field.  Therefore, a prediction similar to that
for the field $a$ exists: the quartic
terms $d^4$ induced by the cubic interactions (to all levels)
must be cancelled by the elementary quartic interaction
of four dilatons.  We will verify this prediction.

In closed string field theory all quartic terms that have been
computed to date involve states that are primary and have $c \bar c$
ghost dependence.  These states
are off-shell only because their dimension is not zero.
The computations of elementary four-string couplings in this
paper involve dilaton states,
which are nonprimary states with non-standard
ghost dependence.  The antighost insertions of the four-string
interaction become quite nontrivial: they are not
of the form $b_{-1} \bar b_{-1}$ acting on the moving puncture.
The steps that must be taken using Moeller's data to compute
such general four-point interactions are explained in detail
in Section~\ref{sect3}.

Our analysis focuses on the two-dimensional  space of deformations
generated  by $a$ and $d$. The simultaneous marginality implies that the
cancellation between cubic and quartic contributions in the effective
potential holds for $a^4, d^4$ {\em and} $a^2 d^2$.  The computations
of the elementary quartic amplitudes $d^4$ and $a^2d^2$ are done in
Section~\ref{sect4}.  The success in testing these
cancellations provides evidence that
the setup in Section~\ref{sect3} works correctly and that the data of Moeller
captures  sophisticated information about the local coordinates on
the punctures of a class of spheres that enter into the string vertex.
Since the cancellations must happen for any  four-string
vertex that is consistent with the chosen three-string vertex,
the amplitudes that are integrated over $\mathcal{V}_{0,4}$ are in fact
total derivatives.  In this way the quartic couplings only depend on
the boundary $\p\mathcal{V}_{0,4}$, which is  indeed determined by
the three string vertex by the condition of gauge invariance.

Our interest on computations involving the dilaton arises from
  additional reasons.  At some degree of accuracy most closed
string theory computations
involve the dilaton.
Condensation of the dilaton plays a role in the Hagedorn transition:
the coupling of the dilaton to nearly relevant states suggests that this
transition is first order and occurs below
the Hagedorn temperature~\cite{Atick:1988si}. The dilaton must
certainly condense
in the decay of orbifold
cones~\cite{Adams:2001sv,Gregory:2003yb,Headrick:2003yu}.
Finally, we  expect the dilaton to be relevant to the computation of
the bulk tachyon potential.

In  Section~\ref{sect2} we focus on computations that involve only
the quadratic and cubic terms in the closed string field theory action.
We begin by calculating
the effective potential $\mathbb{V}(a,d)$ obtained by integrating
out the tachyon field.
We find that
the domain of definition of the marginal direction $a$ is bounded,
as it was for the Wilson line parameter in open
string field theory~\cite{Sen:2000hx} but interestingly,
at least to this level, dilaton deformations are not bounded. This suggests
the attractive possibility that closed string field theory may be able
to describe arbitrarily large dilaton deformations (additional evidence is
discussed in section 5).
We then compute contributions to the quartic terms in
$\mathbb{V}(a,d)$ from closed
string states of level less than or equal to six. This gives us enough data
to perform a rough extrapolation to infinite level. For the terms quartic
on the dilaton in $\mathbb{V}(a,d)$
we push the calculation to higher level by exploiting the
factorization of
correlators.  Intriguingly, the closed string computation is
related to a computation in the {\em quantum} gauge-fixed open string
action.

In section~\ref{sect3}  we discuss the computation
of general quartic elementary interactions, paying particular attention to
the antighost insertions and collecting a series of results that allow the
straightforward calculation of such interactions.
In section~\ref{sect4} we perform the computations
of the quartic couplings $a^2 d^2$ and $d^4$ needed for $\mathbb{V}(a,d)$.

Section~\ref{sect5} is our concluding section.
We analyze in detail the expected cancellations using an  infinite-level
extrapolation of the cubic contributions. We discuss
a definition of level suitable for quartic interactions and  find evidence
that as the level is increased quartic interactions are suppressed,
just as it happens for cubic
interactions. Finally, we state some open problems and suggest possible
directions for investigation.

\sectiono{Dilaton and marginal field potential from cubic interactions}
\label{sect2}

With $\alpha'=2$ the closed string field potential $V$ is
given by~\cite{Zwiebach:1992ie,Saadi:tb}
\begin{equation}
\label{csftpot}
\kappa^2 V=\frac{1}{2}\langle \Psi|c_0^-\, Q|\Psi\rangle
+\frac{1}{3!} \{ \Psi,\Psi,\Psi\} +\frac{1}{4!} \{
\Psi,\Psi,\Psi,\Psi\} +\cdots\,\,.
\end{equation}
A state $|\Psi\rangle$ in closed string spectrum is a
ghost number two state that satisfies
$(L_0-\bar L_0)|\Psi\rangle =0$ and $(b_0-\bar b_0)|\Psi\rangle
=0$. We fix the gauge invariance of the theory using the
Siegel gauge $(b_0+\bar b_0)|\Psi\rangle =0\,.$
The level $\ell$ of a state is defined by
$\ell =L_0+\bar L_0+2\,.$
The closed string tachyon has level zero and fields corresponding
to marginal directions have level two.
We have $c_0^\pm = {1\over 2} (c_0 \pm \bar c_0)$ and
the BRST operator is $Q= c_0L_0 + \bar c_0 \bar L_0 + \dots$, where
the dots denote terms independent of $c_0$ and of $\bar c_0$.
We normalize correlators using $ \langle 0| c_{-1}\bar c_{-1} c_0^- c_0^+ c_1
\bar c_1|0\rangle=1$ and note that
  \begin{equation}
\label{faccorr}
\langle  c(z_1)c(z_2)c(z_3) \,   \bar c (\bar w_1)\bar c (\bar
w_2)\bar c (\bar w_3)
\rangle = -2 \langle  c(z_1)c(z_2)c(z_3) \rangle_o \,\cdot\,\langle
\bar c (\bar w_1)\bar c (\bar w_2)\bar c (\bar w_3) \rangle_o\,,
\end{equation}
where
$\langle  c(z_1)c(z_2)c(z_3) \rangle_o = (z_1-z_2)(z_1-z_3) (z_2 - z_3)$
is the conventional open string field theory  correlator.
The cubic amplitude for three zero-momentum closed string tachyons is
\begin{equation}
\langle c_1\bar c_1,c_1\bar c_1,c_1\bar
c_1\rangle=  2 \mathcal{R}^6\,, \quad \hbox{with} \quad
\mathcal{R} \equiv {1\over
\rho} = {3\sqrt{3}\over 4} \simeq 1.2990\,.
\end{equation}
Here $\rho$ is the (common) mapping radius of the disks that define the
three-string vertex.

\subsection{Direct closed string computation}\label{dcdsdlv}

In our previous work~\cite{Yang:2005iu} we calculated the potential
for the marginal direction $a$ associated with the
state $\alpha_{-1}\bar \alpha_{-1}\, c_1\bar
c_1|0\rangle$.  This time we want
to include the zero-momentum ghost dilaton $d$
associated with the state $(c_1c_{-1}-\bar c_1\bar c_{-1})|0\rangle$.
This state does not  fit the
strict definition of a marginal state: it has dimension $(0, 0)$ but
it is not a Virasoro primary. Nevertheless, the dilaton theorem indicates
that this field has a vanishing potential.
Including the tachyon and the massless fields the string field is
\begin{equation}
\label{thefirstexp}
|\Psi_0\rangle=t\, c_1\bar
c_1|0\rangle + a \,\alpha_{-1}\bar \alpha_{-1}\, c_1\bar
c_1|0\rangle +d (c_1c_{-1}-\bar c_1\bar c_{-1})|0\rangle.
\end{equation}
The subscript on the string field (and in the potentials below) denotes
the level of the highest level {\em massive} field included.
  The ghost structure of the zero-momentum dilaton implies
that the cubic vertex cannot couple  {\it one} dilaton to any quadratic
combination of $t$ and $a$.  Moreover, the cubic vertex
cannot couple  three dilatons, nor it can couple two dilatons  to an $a$.
The only possible three-point coupling that involves a dilaton is
$t d^2$.  The corresponding  term in the potential $\kappa^2 V$  is
\begin{equation}
{1\over 3!}\cdot 3 \cdot (-2)\cdot   t\,d^2  \langle c_1 \bar c_1\, ,
  c_{-1} c_{1}\, ,  \bar c_{-1} \bar c_{1} \rangle ,
\end{equation}
where the factor of 3 arises from three possible ways to choose the puncture
for the tachyon and the $(-2)$ from the two cross-terms that contribute from
the dilaton insertions. Using the
factorization property (\ref{faccorr}), we find
\begin{equation}
  +2 \, t\,d^2
\langle c_1\,,
c_{-1} c_{1}\,, 1\rangle_o\,\langle \bar c_1\,,1\, ,\bar
c_{-1}\bar c_{1}\rangle_o =  2\, t\,d^2  \Bigl(-{8\over 27}
\mathcal{R}^{3}\Bigr)
\Bigl( {8\over 27}\mathcal{R}^{3}\Bigr)  = -\frac{27}{32}\, t\, d^2\,.
\end{equation}
To evaluate the open string amplitudes we have used the conservation law for
the $c_{-1}$ ghost oscillator (see \cite{Rastelli:2000iu}, eqn.~(4.6)).
The potential computed without including vertices higher than
cubic~is:
\begin{equation}
\label{pot24} \kappa^2 V_{(0)}=-t^2+\frac{6561}{4096}
\, t^3+\frac{27}{16}\,t a^2-\frac{27}{32}\, t d^2.
\end{equation}
To find the effective potential $V(a,d)$ we solve
for $t$ as a function of $a$ and $d$. One readily finds:
\begin{equation}
\label{tvm}
t^{V/M}= {8\over 19683} \,\bigl(\, 512 \pm
\sqrt{2}\sqrt{131\hskip0.5pt 072-1062\hskip0.5pt 882
\,a^2+531\hskip0.5pt 441\,
d^2}\,\bigr) \,.
\end{equation}
There are two branches: the vacuum ($V$) branch and the marginal ($M$) branch.
In the vacuum branch the tachyon has finite expectation value when $a$ and
$d$ vanish -- the expectation value corresponding to the stationary point of
the cubic potential.  In the marginal branch the tachyon expectation value
vanishes when $a$ and $d$ vanish.

It follows from (\ref{tvm}) that the effective potential
is well defined as long as
\begin{equation}
\label{hyp}
1062\hskip0.5pt 882
\,a^2\, - \, 531\hskip0.5pt 441\,
d^2  \leq 131\hskip0.5pt 072  \,.
\end{equation}
This implies that for $d=0$ we must have $|a|\leq {256\over 729}
\simeq 0.3512$.
For $a=0$, however, there is no constraint on the magnitude of $d$!
This is a consequence
of the minus sign with which the term $t\,d^2$ appears in the
potential (\ref{pot24});
the tachyon couples with {\em opposite} signs to $a^2$ and $d^2$.
This is our first
indication that closed string field theory is able to describe
arbitrarily large dilaton
deformations (in Section~\ref{sect5} we cite further evidence to this effect).

\begin{figure}
\centerline{\hbox{\epsfig{figure=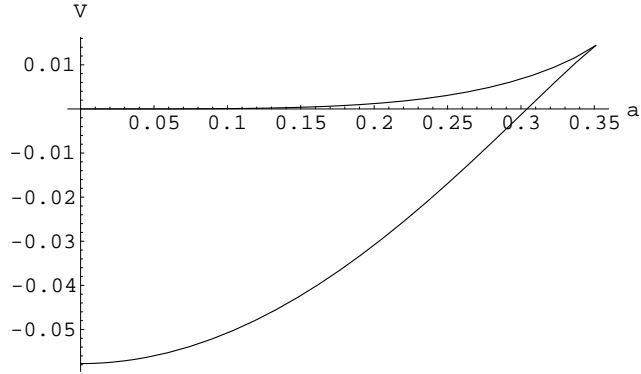,
height=5cm}}} \caption{The effective potential
for the radius deformation in the marginal branch
(top) and in the tachyon branch (bottom).} \label{amarginal}
\end{figure}

\begin{figure}
\centerline{\hbox{\epsfig{figure=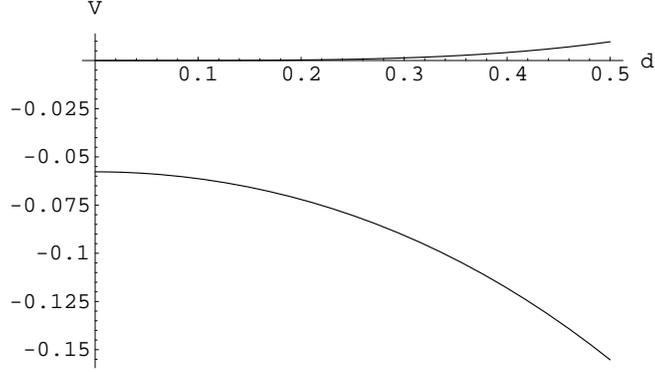,
height=5cm}}} \caption{The effective potential
for the dilaton in the marginal branch (top), and
in the vacuum branch (bottom).} \label{dmarginal}
\end{figure}

In Figure~\ref{amarginal} we show the effective potential in the
$a$-direction $(d=0)$
in both the marginal and tachyon branches.  The qualitative features
of this potential
match those of the potential for the marginal deformation $c\partial X$ in
open string field
theory~\cite{Sen:2000hx}:  the marginal branch is reasonably flat and the
two branches meet at the maximum possible value for $a$.
In Figure~\ref{dmarginal} we show the effective potential in the
$d$-direction $(a=0)$
in both the marginal and tachyon branches.  The marginal branch is
roughly flat and the
vacuum branch curves downward; the two branches do not meet.
$a$ and zero $d$).

Since the status of the  stationary point in the
cubic tachyon potential is still unclear, we focus henceforth
on the marginal branch.  We aim to use the quadratic and
cubic terms in the string field theory to calculate
  the terms in  $V(a,d)$ that are quartic in $a$ and $d$,  the leading
terms for small $a$ and $d$. We then want to show that these terms
are cancelled by elementary quartic interactions.
In order to calculate quartic terms, the $t^3$ interaction in
(\ref{pot24}) is not needed: the tachyon $t$ is at least quadratic
in $a$ and $d$, so this interaction would contribute terms
of order six in the marginal fields.  Solving for the tachyon $t$
as a function of $a$ and $d$ and substituting back in the potential,
the quartic
terms $\mathbb{V}_{(0)}$ in our  calculation are:
\begin{equation}
\label{24result}
\kappa^2 \mathbb{V}_{(0)}  = {3^6\over 2^{10} }\Bigl( a^4 - a^2 d^2
+ {d^4\over
4} \Bigr)
\simeq    0.7119~ a^4 -  0.7119~a^2 d^2 + 0.1780 \, d^4\,.
\end{equation}

We now turn to the computation
  to higher level, still using only the cubic vertex of
the theory. In order to generate the required string field
we note that the states are built with
  oscillators $\alpha_{n\leq -1}, \bar\alpha_{n\leq -1}$
of the coordinate $X$,
Virasoro operators $L'_{m\leq -2}, \bar L'_{m\leq -2}$
corresponding to the remaining coordinates (thus $c=25$),
and ghost/antighost oscillators.
We can list such fields systematically using the generating
function:
\begin{eqnarray}
\label{generating function}
f(x,\bar x,y, \bar y)&=&
\prod_{n=1}^\infty \frac{1}{1-\alpha_{-n} x^n}\,
\frac{1}{1-\bar\alpha_{-n} \bar x^n}\,\prod_{m=2}^\infty
\frac{1}{1-L'_{-m} x^m} \,\frac{1}{1-\bar L'_{-m}
\bar x^m}\nonumber\\
&&
\cdot\prod_{k=-1\atop k\not=0}^ \infty   (1 + c_{-k} x^k y) (1+\bar
c_{-k} \bar x^k
\bar y)
\prod_{l=2}^\infty (1+b_{-l} x^l y^{-1})\, (1+\bar b_{-l}
\bar x^l \bar y^{-1})\,.
\end{eqnarray}
A term of the form  $x^{n} \bar x^{\bar n}  y^m \bar y^{\bar m}$
corresponds to a state with $(L_0, \bar L_0) = (n,\bar n)$ and
ghost numbers $(G, \bar G )= (m,\bar m)$.  Since the string
field must have total ghost number two we require $m+ \bar m = 2$.
A massive field $M$ is relevant to our calculation  if it has a
coupling $M a^2$, or $Md^2$, or $Mad$, or any combination of them.
If all three couplings vanish we can set $M$ to zero consistently
to this order.  We readily see the following rules also apply:

\begin{itemize}

\item A field with $(G, \bar G) = (1,1)$ can couple only to $a^2$
and to $d^2$. Such field must have an even number of $\alpha$'s
and an even number of $\bar \alpha$'s.

\item  A field with $(G, \bar G) = (0,2)$ or $(2,0)$ can couple
only to $a d$. Such field must have an odd number of $\alpha$'s
and an odd number of $\bar \alpha$'s.

\item  A field with $(G, \bar G) = (-1,3)$ or $(3,-1)$ can couple
only to $d^2$. Such field must have an even number of $\alpha$'s
and an even number of $\bar \alpha$'s.

\end{itemize}

At level $\ell =4$ we have $L_0=\bar L_0=1$.  The coefficients of $(x\bar x
y\bar y, x\bar x y^2, x\bar x \bar y^2) $ give all possible  terms in
the string field. With the above rules the  set is reduced to
\begin{equation}
\label{lev4sfml}
\begin{split}
|\Psi_4\rangle &=\phantom{+}f_1\, c_{-1}\bar
c_{-1}+\,f_2\, L'_{-2}\bar
L'_{-2} c_1 \bar
c_1 +(\,f_3\, L'_{-2} c_1\bar c_{-1}+ f_4\,
\bar L'_{-2} c_{-1}\bar c_1)  \\[0.5ex]
&+\,r_1\,
\alpha^2_{-1}\bar\alpha^2_{-1} c_1\bar c_1
+ (\,r_2\,\alpha_{-1}^2 c_1\bar c_{-1}+ r_3\,
\bar\alpha_{-1}^2 c_{-1} \bar c_1) \\[0.5ex]
&+\,r_4\,\alpha^2_{-1}\bar L'_{-2} c_1\bar c_1+
        r_5\,L'_{-2}\bar \alpha^2_{-1}  c_1\bar
c_1
+ (\,r_6\,\alpha_{-1}\bar\alpha_{-1} c_{-1}c_1+
        r_7\,\alpha_{-1}\bar\alpha_{-1} \bar c_{-1}\bar c_1)\,.
\end{split}
\end{equation}
The corresponding terms in the potential $V_{(4)}$ are given in
Appendix~\ref{a1}.
Eliminating all massive fields through their equations of motion
we obtain
\begin{eqnarray}
\label{dfjgkjeku}
\kappa^2 V_{(4)} &=& -\frac{19321}{46656}~ a^4 +\frac{1619}{15552}
~a^2 d^2-\frac{6241}{186624}~d^4 \nonumber \\[0.9ex]
&\simeq &-0.4141\, a^4+0.1041\, a^2 d^2-0.0334\,  d^4\,.
\end{eqnarray}
To get the total contribution up to level four we  add the
above to the  result in (\ref{24result}):
\begin{eqnarray}
\label{tot48}
\kappa^2 \mathbb{V}_{(4)} &=&\frac{222305}{746496}~ a^4 -\frac{151243}{248832}
~a^2 d^2+\frac{431585}{2985984}~d^4 \nonumber \\[0.9ex]
&\simeq &0.2978\, a^4-0.6078\, a^2 d^2+0.1445\,  d^4\,.
\end{eqnarray}

We have computed  the contribution from level six string fields.
The states that contribute as well as the potential
are given in Appendix~\ref{a1}.
Eliminating out the massive fields we find:
\begin{eqnarray}
\label{98htfgkuh}
\kappa^2 V_{(6)}&=&\frac{53824}{531441} \,a^2 d^2- \frac{5000}{177147}\, d^4
\nonumber \\[0.5ex]
&\simeq&
0.1013\, a^2 d^2 - 0.0282\, d^4.
\end{eqnarray}
The full set of quartic terms to level six is obtained by adding
the above to (\ref{tot48}):
\begin{eqnarray}
\kappa^2 \mathbb{V}_{(6)}&=& \frac{222305}{746496} \,a^4-\frac{275652665}
{544195584}\, a^2 d^2+\frac{84395155}{725594112}\, d^4 \nonumber\\[1.0 ex]
&=&0.2978\, a^4-0.5065\, a^2 d^2+0.1163\, d^4\,.
\end{eqnarray}
By comparing the successive approximations $\mathbb{V}_{(0)}$,
$\mathbb{V}_{(4)}$, and $\mathbb{V}_{(6)}$ we note that the
coefficients of $a^4, a^2d^2,$ and $d^4$ all decrease (in magnitude)
as we increase the level.  Introducing the notation
\begin{equation}
\label{notresults}
\begin{split}
\kappa^2 {V}_{(\ell)} &=  {c}_{a^4}(\ell)  \, a^4
+ {c}_{a^2d^2}(\ell) \,a^2\, d^2  + c_{d^4}(\ell) \, d^4 \\
\kappa^2 \mathbb{V}_{(\ell )} &=  \mathcal{C}_{a^4}(\ell)  \, a^4
+ \mathcal{C}_{a^2d^2}(\ell) \,a^2\, d^2  + \mathcal{C}_{d^4}(\ell) \, d^4
\end{split}
\end{equation}
the  information obtained so far is collected in
Table~\ref{cubic_table}. The table shows both the contributions
that arise from each level $\ell$ of the massive fields
(the quantities $c_{\dots}(\ell)$)
and the total contributions up to that level (the
quantities $\mathcal{C}_{\dots}(\ell)$). Even at infinite level the
total contributions to the quartic coefficients
do not vanish.  The infinite-level cubic calculations give
a quartic potential that must be cancelled by
the contributions from the elementary quartic interactions.

\begin{table}[ht]
\begin{center}
{\renewcommand\arraystretch{1.1}
\begin{tabular}{|c|c|c|c|c|c|c|}
            \hline
            $~~\ell$
            & $c_{a^4}(\ell)$
            &~~ $\mathcal{C}_{a^4}(\ell)$~~
            & $c_{a^2 d^2}(\ell)$
            & $\mathcal{C}_{a^2 d^2}(\ell)$
            & $c_{d^4}(\ell)$
            &~~ $\mathcal{C}_{d^4}(\ell)$~~  \\[0.1ex]
            \hline
            $0$ & $\phantom{-}0.7119$
            & $0.7119$ & $-0.7119$
            & $-0.7119$& $\phantom{-}0.1780$ & 0.1780 \\
            \hline
            $4$  &$-0.4141$ & $0.2978$
            & $\phantom{-}0.1041$ & $-0.6078$& $-0.0334$ & $0.1445$ \\
            \hline
            $6$  &\phantom{-}0 & $0.2978$
            & $\phantom{-}0.1013$ & $-0.5065$& $-0.0282$ & $0.1163$\\
            \hline
\end{tabular}}
\end{center}
\caption{\small The coefficients of the quartic terms in the
effective potential
for $a$ and $d$ as a function of the level $\ell$ of the massive fields
integrated out.
  } \label{cubic_table}
\end{table}

\subsection{Contributions to $d^4$ calculated using OSFT}

To extend the computations of the previous
subsection to higher levels requires significant work. In~\cite{Yang:2005iu},
we showed how to obtain
the contributions to $a^4$ in the closed string effective potential
in terms of analogous contributions to the potential of the Wilson
line parameter $a_s$ in open string field theory.
This comparison worked because, in the Siegel gauge, the closed
string kinetic and cubic terms factorize into holomorphic and
antiholomorphic correlators which feature in analogous open
string computations.

In this subsection we compute additional
coefficients $c_{d^4}(2\ell)$ in the dilaton potential.  The factorization
property cuts down significantly the number of
correlators that must be computed.  As we show at the end of this subsection,
the computation of the dilaton potential is at least formally
related to a computation in the  {\em quantum}
  gauge fixed open string field theory.

   The three string vertex
couples $d^2$ to massive fields  of ghost number $(1,1)$, $(-1,3)$, and
$(3,-1)$. We thus consider the massive closed string field of level $2\ell$:
\begin{equation}
\label{masssf4eeiuftfh} |\Psi\rangle
=\sum_{i,j}\psi_{ij}\,|\mathcal{O}_i\rangle\otimes
|\overline{\mathcal{O}}_j \rangle + \sum_{\alpha, a}
\bigl(\psi_{\alpha a}\, |\mathcal{O}_\alpha\rangle\otimes
|\overline{\mathcal{O}}_a\rangle\, +\, \psi_{a\alpha}\,
|\mathcal{O}_a\rangle \otimes
|\overline{\mathcal{O}}_\alpha\rangle\bigr)\,.
\end{equation}
The holomorphic basis states $|\mathcal{O}_i\rangle,
|\mathcal{O}_\alpha\rangle$, and $|\mathcal{O}_a\rangle$ are all
level $\ell$ open string states in the Siegel gauge, with ghost
numbers $1,-1$, and $3$, respectively:
\begin{equation}
G(\mathcal{O}_i)=1\,, \quad G(\mathcal{O}_\alpha)=-1\,, \quad
   G(\mathcal{O}_a)=3 \,.
\end{equation}
The barred states are identical looking states built with antiholomorphic
oscillators.

The computation of the quadratic and cubic terms in the closed string
action is helped by the following definitions:
\begin{equation}
\label{11def}
\begin{split}
& m_{ij}\equiv\langle
\mathcal{O}_i|c_0|\mathcal{O}_j\rangle=m_{ji}\,, \quad
R_{\alpha a} \equiv \langle  \mathcal{O}_\alpha|c_0|\mathcal{O}_a
\rangle = \langle
\mathcal{O}_a|c_0|\mathcal{O}_\alpha
\rangle\,, \\[1.0ex]
& K_i\equiv \langle \mathcal{O}_i,c_1c_{-1},\mathbf{1}\rangle\,
,\quad P_a\equiv \langle
\mathcal{O}_a,\mathbf{1},\mathbf{1}\rangle\, ,\quad Q_\alpha
\equiv \langle  \mathcal{O}_\alpha ,c_1c_{-1},c_1c_{-1} \rangle\,.
\end{split}
\end{equation}
We now want to evaluate the potential
\begin{equation}
\label{csftpot} \kappa^2 V_{2\ell}=\frac{1}{2}\langle \Psi|c_0^-\,
Q_B|\Psi\rangle +\frac{1}{2} \{ \Psi,D,D\} \,\,,
\end{equation}
where $|D\rangle =d\, (c_1c_{-1} - \bar c_1 \bar c_{-1} )
|0\rangle $ and the state $|\Psi\rangle$ is given in
(\ref{masssf4eeiuftfh}). Using the factorization property of
correlators and the above definitions we find
\begin{equation}
\begin{split}
\kappa^2 V_{2\ell} &=(\ell-1)\, m_{ii'}\, m_{jj'}\,
\psi_{ij} \, \psi_{i'j'}
+ 2 (\ell-1)R_{\alpha b}\,R_{\beta a}\, \psi_{\alpha a}\, \psi_{b\beta}\,
   \\[0.7 ex]
&~~~ +2(-)^\ell\,  K_i\, K_j\, \psi_{ij}\, d^2\,
\, - Q_\alpha\,
P_a\, \psi_{\alpha a}\, d^2- P_b\, Q_\beta\, \psi_{b\beta}\,
d^2.
\end{split}
\end{equation}
where repeated indices are summed over. The factor
$(-1)^\ell$ in the third term of the right hand side
arises because $\langle \mathcal{O}_i, \mathbf{1}, c_1 c_{-1}
\rangle=\Omega_{\mathcal{O}_i} \langle \mathcal{O}_i, c_1
c_{-1},\mathbf{1} \rangle=(-1)^\ell K_i$. Solving the (linear)
equations of motion for $\psi_{ij}, \psi_{a\alpha},$ and
$\psi_{\alpha a}$ and substituting back into the potential one
finds
\begin{equation}
\label{clkfkjbdf}
\kappa^2 V_{2\ell} = c_{d^4}(2\ell) \, d^4 \,, \quad
\hbox{with} \quad c_{d^4}(2\ell)=-\,\frac{1}{\ell-1}\Big[\Big(K^T\, M^{-1}\,
K\Big)^2+\frac{1}{2}\Big(P^T\, R^{-1}\, Q\Big)^2\Big].
\end{equation}
The first term in the bracket gives the contribution to the potential
from the $(1,1)$ massive fields and the second term gives the
contribution to the potential from the $(-1,3)$ and $(3,-1)$ massive fields.

Let us now compute $c_{d^4}$ for levels ranging from
zero to ten. We use the open string universal
subspace with matter Virasoro operators of central charge $c=26$.
The twist property
\begin{equation}
\langle A,B,C\rangle=
\Omega_A\Omega_B\Omega_C (-1)^{BC+1} \langle A, C,B\rangle\,
\end{equation}
of the cubic open string vertex
implies that for $B$ Grassmann even $\langle A,B,B\rangle = -\Omega_A
\langle A,B,B\rangle$. Consequently $\langle A,B,B\rangle$
vanishes for twist even $A$ or, equivalently,
for states $A$ of even level.
We deduce that the vectors $P_a$ and $Q_\alpha$
vanish for states at even levels.

\begin{itemize}
\item At $\ell=0$ the only massive state is the tachyon:
\begin{eqnarray}
M=1,\hspace{5mm} K=\frac{3\sqrt 3}{8}, \hspace{5mm}   \Rightarrow
c_{d^4}(2)=\frac{729}{4096}\simeq 0.177979,
\end{eqnarray}
which confirms our result in (\ref{24result}).

\item At $\ell=2$ there are two ghost number one states,
$c_{-1}|0\rangle$ and $L_{-2} c_1|0\rangle$. We find
\begin{eqnarray}
M&=&\hbox{diag}(-1,~13), \hspace{5mm} K=\left(
\begin{array}{cc}
    \frac{\sqrt 3}{24}, & -\frac{65}{24\sqrt 3} \\
\end{array}
\right),\hspace{5mm} K^T\, M^{-1}\, K=\frac{79}{432}\,,\nonumber\\[0.5 ex]
&\Rightarrow& c_{d^4}(4)=-\frac{6241}{186624}\simeq -0.0334416,
\end{eqnarray}
which  agrees with the result in (\ref{dfjgkjeku}).

\item At $\ell=3$ there are three ghost number one states:
$c_{-2}|0\rangle$, $b_{-2}c_{-1}c_1|0\rangle$, and $L_{-3}
c_1|0\rangle$; one ghost number minus one state $b_{-2}|0\rangle$ and
one ghost number three state $c_{-2}c_{-1}c_1|0\rangle$. Using the order
in which we listed the states the
relevant matrices are:
\begin{eqnarray}
M&=&\left(
\begin{array}{ccc}
    0 & -1 & 0   \\
    -1 & 0 & 0   \\
    0 & 0 & -52   \\
\end{array}
\right), \hspace{5mm} K=-\frac{5}{27}\left(
\begin{array}{c}
    1 \\
    2 \\
    0 \\
\end{array}
\right)~\Rightarrow ~K^T M^{-1} K=-\frac{100} {729}\,,
\nonumber\\[1.0ex]
R&=&1\,,\hspace{5mm} Q= -\frac{20}{27}\,,\hspace{5mm}
P=-\frac{10}{27}\,,\quad \Rightarrow \quad P^T\, R^{-1}\, Q=\frac{200}{729}\,.
\end{eqnarray}
The contribution to $c_{d^4}$ is then
\begin{equation}
c_{d^4}(6)
=-\frac{5000}{531441}-\frac{10000}{531441}
=-\frac{5000}{177147}\simeq -0.0282251,
\end{equation}
which is the result obtained in (\ref{98htfgkuh}).

\item At level $\ell=4$, $P= Q=0$ as we argued before. There are
six ghost number one states:
\begin{equation}
c_{-3}|0\rangle\,, \quad b_{-3}c_{-1} c_1|0\rangle,\quad
b_{-2}c_{-2}
c_1|0\rangle,  \hspace{5mm}
L_{-2} c_{-1}|0\rangle, \hspace{5mm}
L_{-2}^2 c_1|0\rangle,
\hspace{5mm}  L_{-4}c_1|0\rangle\,.
\end{equation}
We find  $M = \begin{pmatrix}
0&1\cr 1&0\end{pmatrix}\oplus~  \hbox{diag} (\, -1,\,-13\,,\,
442\,,\, 130\,)$ as well
as
\begin{equation}
K^T=\Bigl(
-\frac{5}{27\sqrt 3}\,, \,
   \frac{5}{9\sqrt 3}\,, \,
\, -\frac{343}{648\sqrt 3}\,,\,
   -\frac{65}{216\sqrt  3}\,,
\, \frac{3523}{324\sqrt 3}\,,\,
   \frac{65}{12\sqrt 3}\,\Bigr).
\end{equation}
We find  $K^T M^{-1} K = {37\over 396576}$, and therefore,
\begin{eqnarray}
c_{d^4}(8)=\,-\frac{1369}{471817571328}\simeq
-2.9\times 10^{-9}.
\end{eqnarray}
This number is anomalously small.  Clearly, the coefficients
$c_{d^4}(\ell)$ do not settle into a regular pattern for small
$\ell$.

\item At level $\ell=5$, there are nine ghost number one states:
\begin{eqnarray}
&&c_{-4}|0\rangle, \hspace{5mm} b_{-4}c_{-1}c_1|0\rangle,
\hspace{5mm} b_{-3}c_{-2} c_1|0\rangle, \hspace{5mm} b_{-2}c_{-3}
c_1|0\rangle, \hspace{5mm}  L_{-2}c_{-2} |0\rangle, \hspace{5mm}
L_{-2} b_{-2}c_{-1} c_1|0\rangle,\nonumber\\
&& L_{-3} c_{-1}|0\rangle, \hspace{5mm} L_{-3}L_{-2} c_1|0\rangle,
\hspace{5mm} L_{-5}c_1|0\rangle,
\end{eqnarray}
three ghost number minus one states, and three ghost number three states:
\begin{eqnarray}
&&\mathcal{O}_\alpha: \hspace{10mm} b_{-4}|0\rangle,\hspace{18mm}
b_{-3}b_{-2} c_1|0\rangle,\hspace{10mm} b_{-2}L_{-2}
|0\rangle,\hspace{10mm} \nonumber\\
&&\mathcal{O}_a:
\hspace{10mm}c_{-4}c_{-1}c_1|0\rangle,\hspace{10mm}
c_{-3}c_{-2}c_1|0\rangle,\hspace{10mm}
   c_{-2}c_{-1}c_1 L_{-2}
|0\rangle\,.
\end{eqnarray}
We then find:
\begin{eqnarray} M&=&\left(
\begin{array}{cc}
   0 & -1 \\
   -1 & 0 \\
\end{array}
\right)\oplus \left(
\begin{array}{cc}
   0 & 1 \\
   1 & 0 \\
\end{array}
\right)\oplus \left(
\begin{array}{cc}
   0 & -13 \\
   -13 & 0 \\
\end{array}
\right)\oplus \hbox{diag}(52, -936, -260), \nonumber\\[1.0ex]
K^T&=&\frac{1}{729}\left(
\begin{array}{ccccccccc}
    39, & 156, & 75, & -50,& 325, & 650, & 0, & 0,  & 0 \\
\end{array}
\right),
\end{eqnarray}
and
\begin{eqnarray}
R=\hbox{diag}(1,~1,~13),\hspace{5mm} Q^T=\left(
\begin{array}{ccc}
    \frac{104}{243}, & \frac{100}{243}, & \frac{1300}{729} \\
\end{array}
\right), \hspace{5mm} P^T=\left(
\begin{array}{ccc}
     \frac{26}{243}, & \frac{50}{729}, & \frac{650}{729} \\
\end{array}
\right).
\end{eqnarray}
Therefore, one obtains:
\begin{eqnarray}
K^T M^{-1} K&=& -\frac{52168}{531441},\hspace{7mm} PR^{-1}Q=
\frac{104336}{531441}.\nonumber\\[1.5 ex]
c_{d^4}(10)&=& c_{d^4}^{(1,1)} (10)+ c_{d^4}^{(-1,3)}(10)=
-\frac{680375056}
{282429536481}(1+2)\nonumber\\[0.5 ex]
&=&-\frac{680375056} {94143178827}\simeq -0.007227\,.
\end{eqnarray}
The total contribution up to level $\ell=10$ is thus:
\begin{eqnarray}
\mathcal{C}_{d^4}(\ell=10)=\frac{12156561955612607}
{111441423077388288}\simeq 0.109085.
\end{eqnarray}
\end{itemize}

\begin{table}[ht]
\begin{center}
{\renewcommand\arraystretch{1.1}
\begin{tabular}{|c|c|c|}
             \hline
             $~~\ell~~$
             &  $c_{d^4}(\ell)$
             &~~ $\mathcal{C}_{d^4}(\ell)$~~  \\[0.1ex]
             \hline
             $0$ & $\phantom{-}0.1780$ & 0.1780 \\
             \hline
             $4$  & $-0.0334$ & $0.1445$ \\
             \hline
             $6$  & $-0.0282$ & $0.1163$\\
             \hline
             $8$ & $-3\times 10^{-9}$ & $0.1163$\\
             \hline
             $10$ & $-0.0072$ & $ 0.1091$\\
             \hline
\end{tabular}}
\end{center}
\caption{\small The coefficients of the quartic terms in the
effective potential for  $d$ as a function of the level $\ell$ of
the massive fields integrated out.
   } \label{cubic_tablexx}
\end{table}

\noindent
\underline{Speculations on related open string computation.}
The above closed string computation requires holomorphic correlators
that appear naturally in an open string computation.
We consider the gauge-fixed open string field theory and include
spacetime fields $u$ and $v$ of ghost numbers  one and minus one:
\begin{eqnarray}
|A\rangle= |0\rangle u+c_{1} c_{-1}|0\rangle v.
\end{eqnarray}
The spacetime fields appear as coefficients of the ghost number zero
and ghost number two states.  Since the open string field is Grassmann
odd (and the vacuum $|0\rangle$ is Grassmann even) the fields $u$ and
$v$ should be Grassmann odd.  Note that the corresponding states have
$L_0=0$ so they may be viewed as marginals of unusual ghost number.
The kinetic term
that could couple $u$ and $v$ vanishes. In fact, ghost number conservation
implies that terms coupling only $u$ fields or $v$ fields must vanish.
   We claim that
some of the ingredients in the computation of the open string effective
action for $A$ are closely related to those of the dilaton effective
potential.

To calculate the effective potential for $A$ we consider the massive
open string field
\begin{equation}
|\Phi\rangle =\sum_i |\mathcal{O}_i\rangle \phi_i +\sum_\alpha
|\mathcal{O}_\alpha\rangle \phi_\alpha + \sum_a |\mathcal{O}_a\rangle
\phi_a,
\end{equation}
and compute the potential
\begin{equation}
g^2 V_\ell =\frac{1}{2}\langle \Phi|\,
Q_B|\Phi\rangle + \langle  \Phi,A,A\rangle  \,\,.
\end{equation}
In this computation the cubic term couples $uv$ to
massive fields, but since
$u^2=v^2=0$ no other couplings appear.  In order to make the computation
  more interesting and have $P_a$ and $Q_\alpha$ feature in the result,
we  assume $uv = - vu$ but, in a departure from the conventional
interpretation,  no longer take $u^2$ and $v^2 $ to vanish (one may imagine
that this is  a computation in which the string fields
are two-by-two matrices and we take $u=\sigma_1$ and $v=\sigma_2$).
Evaluation of the above action then gives:
\begin{eqnarray}
g^2 V_\ell=\frac{1}{2} (\ell-1)\phi_i m_{ii'} \phi_{i'} +(\ell-1)
\phi_\alpha R_{\alpha a}\phi_a+(1+(-1)^\ell) K_i \phi_i (vu)+P_a
\Phi_a u^2 + Q_\alpha \phi_\alpha v^2.
\end{eqnarray}
We see that all the matrices and vectors introduced to write
the closed string effective action appear here.
With the assumption that $u^2$ and $v^2$ are nonvanishing one
can compute the effective potential by integrating out the massive
fields.  The result is
\begin{equation}
\label{adunnnn}
g^2 V_\ell (u,v)=-\frac{1}{\ell-1}\Big(\frac{(1+(-1)^\ell)^2}{2} (K^T\,
M^{-1}\, K)-(P^T\, R^{-1} Q)\Big) (uv)^2.
\end{equation}
In the standard interpretation this potential vanishes
simply because $(uv)^2=0$, with no computation necessary.
In that sense marginality seems preserved.
Still the formal resemblance of (\ref{adunnnn})
to (\ref{clkfkjbdf}) is quite intriguing.
The sum $\sum_\ell V_\ell$ does not seem to converge to zero, so we
do not understand the significance of the contributions $V_\ell$.

It may be of interest to examine, after integration of massive fields,
the quantum gauge fixed action at nonzero momentum. Here quartic terms
would survive.
Any exact relation between the classical closed string
field theory action for the dilaton and the quantum, effective,
gauge-fixed  open string action may be quite
illuminating.

\sectiono{Setting up elementary quartic computations}\label{sect3}

In a very useful piece of work Moeller~\cite{Moeller:2004yy} calculated
the quadratic differential that defines the local
coordinates on the punctures of the four-punctured
spheres that comprise the quartic string vertex.
He also gave a concrete description of the region
of integration $\mathcal{V}_{0,4}$.  This information
is all that is needed, in principle, to compute any
four-string coupling.  In this section we show how to
use this information to set up the computation of
quartic interactions where the anti-ghost insertions
play a nontrivial role -- this happens whenever the
states do not have the simple $c \bar c$ ghost dependence.
The results that we obtain make the computation
of four-string couplings straightforward.  The specific
computations required in this paper will be done
in the following section.

\subsection{The quartic term in the string action}

The description of four-string amplitudes uses the
definition~\cite{Zwiebach:1992ie,Belopolsky:1994sk}
\begin{equation}
\label{multifour}
\{ \Psi_1, \Psi_2, \Psi_3, \Psi_4\} \equiv
{i\over 2\pi} \int_{\mathcal{V}_{0,4}} \hskip-3pt d\lambda_1\wedge
d\lambda_2  \, \langle \Sigma|\, b(v_{\lambda_1}) b(v_{\lambda_2})
|\Psi_1\rangle |\Psi_2\rangle|\Psi_3\rangle|\Psi_4\rangle\,.
\end{equation}
Here $\langle \Sigma|$ is the (operator formalism) surface state
corresponding to a four-punctured sphere $\Sigma \in \mathcal{V}_{0,4}$.
In addition,  $\lambda_1$ and $\lambda_2$ are two real parameters that
describe the moduli space and  $b(v_{\lambda_i})$, with $i=1,2$,
are antighost factors given by
\begin{equation}
b(v_{\lambda_i}) = \sum_{I=1}^4 \sum_{m=-1}^\infty \bigl( B_{i,m}^I b_m^I +
\overline{B_{i,m}^I} \,\bar  b_m^I \bigr) \,,\quad\hbox{with}\quad
B^I_{i,m} = \oint {dw\over 2\pi i} \, {1\over w^{m+2}}  {1\over h'_I}
{\partial h_I\over \partial\lambda_i}\,.
\end{equation}
The functions $h_I (w; \lambda_1, \lambda_2)$, with $I=1, \ldots ,
4$, define maps from a local coordinate $w$ into the sphere
described with uniformizer $z$,  and the primes denote derivatives
with respect to $w$. An overline on an number denotes complex
conjugation. We use a presentation of the moduli space in which
$(\lambda_1,\lambda_2)= (x,y)$, where $\xi = x + iy$ is the
position of the moving puncture. It is then possible to rewrite
the relevant two-form as~\cite{Belopolsky:1994sk}
\begin{equation}
\label{twoforms}
d\lambda_1\wedge d\lambda_2  \,
b(v_{\lambda_1}) b(v_{\lambda_2})  = d\xi\wedge d\bar\xi \,\,
\mathcal{B} \,  \mathcal{B}^\star
\,,
\end{equation}
where
\begin{equation}
\label{exptwoforms}
\mathcal{B} = \sum_{I=1}^4 \sum_{m=-1}^\infty
(B^{I}_m\, b^J_m + \overline{ \Ci^I_m }\, {\bar b}^I_m) \,, \quad
\mathcal{B}^\star = \sum_{I=1}^4 \sum_{m=-1}^\infty
(\Ci^I_m\, b^I_m + \overline{B^I_m} \, {\bar b}^I_m)\,,
\end{equation}
and the coefficients $B^I_{m}$ and $\Ci^I_{m}$ are given by
\begin{equation}
\label{calcBandC}
B^J_{m} = \oint {dw\over 2\pi i} \, {1\over w^{m+2}}  {1\over h'_J}
{\partial h_J\over \partial\xi}\,, \qquad
\Ci^J_{m} = \oint {dw\over 2\pi i} \, {1\over w^{m+2}}  {1\over h'_J}
{\partial h_J\over \partial\bar \xi}\,.
\end{equation}
In (\ref{twoforms}) we have introduced a $\star$-conjugation. Acting
on a number $\star$-conjugation is just complex conjugation. Acting
on a product  of ghost oscillators $\star$-conjugation
  reverses their order and turns holomorphic
oscillators into antiholomorphic ones, and viceversa.  Note that this rule
defines an involution and
is consistent with (\ref{exptwoforms}).
Note
also that in
(\ref{calcBandC}) the functions $h_I (w; \xi, \bar \xi)$ are simply
the functions
that describe the local coordinates written in terms of the complex
modulus $\xi$.

Using (\ref{twoforms}) and  $d\xi\wedge d\bar\xi = -2i
\,dx\wedge dy$ the
multilinear function  in (\ref{multifour}) becomes
\begin{equation}
\label{amplitude}
\{ \Psi_1, \Psi_2, \Psi_3, \Psi_4\} \equiv
{1\over \pi} \int_{\mathcal{V}_{0,4}} \hskip-3pt dx\wedge dy  \,
\langle \Sigma | \,\mathcal{B} \, \mathcal{B}^\star
\,|\Psi_1\rangle |\Psi_2\rangle|\Psi_3\rangle|\Psi_4\rangle\,.
\end{equation}
The quartic term in the string field potential (\ref{csftpot})
  is then given by ($\alpha'=2$)
\begin{equation}
\kappa^2 V = {1\over 4!} \, \{ \Psi, \Psi, \Psi, \Psi\}  =  {1\over
4!} \{\, \Psi^4\,
\}\,.
\end{equation}

The maps from local coordinates to the uniformizing coordinate $z$ on
the four-punctured sphere take the form:
\begin{equation}
\label{coordexp} z=h_I(w; \xi, \bar \xi )=z_I(\xi,\bar \xi)
+\rho_I(\xi,\bar \xi)\,  w +\rho_I^2 \, \beta_I(\xi,\bar \xi)\,
w^2+ \rho_I^3 \gamma_I(\xi,\bar \xi) \, w^3 + \rho_I^4
\delta_I(\xi, \bar \xi) \, w^4 + \cdots  \,.
\end{equation}
Here $\rho_I$ is a real, positive, number called the mapping radius.
For convenience, factors of the mapping radius have been included in the
definition of the higher order
coefficients  $\beta_I, \gamma_I,$ and $\delta_I$.
We choose the first three punctures to be at $z=0$,
$z=1$ and $z=\xi$, so
\begin{equation}
z_1 (\xi, \bar \xi) =0\,, \quad z_2 (\xi, \bar \xi) =1\,, \quad \hbox{and}
\quad z_3 (\xi, \bar \xi) = \xi \,.
\end{equation}
While we can use (\ref{coordexp}) for $I=1,2,3$, the fourth puncture is
placed at $z=\infty$ and one must use a coordinate $t=1/z$ which vanishes
at this point.  We thus write
\begin{equation}
t= h_4(w; \xi, \bar\xi) = \rho_4(\xi,\bar \xi)\,  w +\rho_4^2 \,
\beta_4(\xi,\bar \xi)\,  w^2+ \rho_4^3\, \gamma_4(\xi,\bar \xi) \,
w^3  + \rho_4^4 \delta_4 (\xi\,, \bar\xi) \, w^4 + \ldots \,.
\end{equation}
All operators inserted at the fourth puncture must be thought as inserted
at $t=0$. One must then map them to $z = 1/t$ in order to compute correlators
using the global uniformizer $z$.

Equations (\ref{calcBandC}) allow us to express the coefficients
$B_m^I$ and $\Ci_m^I$ in
terms of the expansion coefficients for the coordinates.  With mode
number minus one (the lowest possible one), we find
\begin{equation}
\label{minusonecoeff}
B_{-1}^{J}=\frac{1}{\rho_3}\delta_{3J},\quad
\Ci_{-1}^{J}=0 \,.
\end{equation}
Note that at this level the antighost insertion is supported
only on the moving puncture, and by our choice $z=\xi$, the
insertion is holomorphic.  Since our string fields are annihilated
both by $b_0$ and $\bar b_0$, the coefficients $B^I_0$ and $\Ci^I_0$
are not needed.

The ghost-dilaton state contains ghost oscillators of mode number minus one.
The coefficients $B_1^I$ and $\Ci_1^I$
are thus needed to compute four-point  amplitudes that involve dilatons.
We then find
\begin{equation}
\label{coeffs1}
B_1^{I}=\rho_I\p \beta_I+ \frac
{1}{2}\rho_3\varepsilon_3\delta_{I3}\,,\quad
\Ci_1^{I}=\rho_I\bar\p \beta_I\,, \quad\hbox{with}
\quad \varepsilon_I\equiv 8\, \beta_I^2- 6\gamma_I\,,
\quad \partial \equiv {\partial
\over \partial \xi}\,, \quad \bar\partial \equiv {\partial \over
\partial \bar \xi}\,.
\end{equation}
A similar calculation gives
\begin{equation}
B_2^I=\rho_I^2 \p (\gamma_I -\beta_I^2)
+ \rho_I^2 (-4 \delta_I-2 \varepsilon_I \beta_I+ 8 \beta_I^3)\delta_{3I},
\quad  \Ci_2^I=\rho_I^2 \bar\p (\gamma_I -\beta_I^2)\,.
\end{equation}
The above results  suffice to compute four point amplitudes with states
that contain oscillators $c_{-n}$ and $\bar c_{-n}$ with
$n\leq 2$. Taking note of the vanishing coefficients,
we see that for states in the Siegel gauge the antighost factor $\mathcal{B}$
is given by
\begin{equation}
\label{Bgeneral}
\mathcal{B} = B_{-1}^{3} b_{-1}^{(3)} +
\sum_{I=1}^4
(B^{I}_1\, b^J_1 + \overline{ \Ci^I_1 }\, {\bar b}^I_1)  +
\sum_{I=1}^4
(B^{I}_2\, b^J_2 + \overline{ \Ci^I_2 }\, {\bar b}^I_2) + \ldots  \,.
\end{equation}
In order to proceed further we learn how to obtain
the coordinate expansion coefficients $\beta_I, \gamma_I$, and $\delta_I$
from quadratic differentials.

\subsection{Strebel differential and local coordinates}

Consider a four-punctured
sphere with uniformizer $z$. Place
the first, second, and fourth punctures at $0, 1,$
and $\infty$, respectively, and let the third puncture
be placed at $z=\xi$.
  The collection of four-punctured spheres that comprise
the moduli space $\mathcal{V}_{0,4}$
can be described as the region of the complex
$z$-plane that contains the allowed values of $\xi$.  These
are the spheres that are {\em not} obtained from the Feynman
diagrams built with one propagator and two three-string vertices.

Each surface in
$\mathcal{V}_{0,4}$ is a four-punctured
sphere with some value of $\xi$. On each surface we consider the
Strebel quadratic differential~\cite{Moeller:2004yy}:
\begin{equation}
\label{strebelqd} \varphi = \phi(z) (dz)^2 \,, \quad \phi(z) = -\,
{(z^2 -\xi)^2\over z^2 (z-1)^2 (z-\xi)^2} \,+\,  {a(\xi, \bar
\xi)\over z (z-1) (z-\xi)} \,.
\end{equation}
Here $a(\xi, \bar \xi)$ is a complex function of $\xi$ {\it and}
$\bar \xi$. While $a$ is not holomorphic, we henceforth write it
as $a(\xi)$, for brevity. If $a$ is known, the quadratic differential
$\phi(z)$ is fully determined.
The quadratic differential has second order poles at the punctures
$z=0, 1, \xi,$ and $\infty$. Expanding around these punctures we find
\begin{equation}
\label{thefirstorderpole}
\begin{split}
\phi(z) &= - {1\over z^2} + {1\over z} \Bigl( -2 -{2\over \xi} +
{a\over \xi}\Bigr)
+ \Bigl( -3 + {1\over \xi} (a-2) + {1\over \xi^2} (a-3) \Bigr)
+ \mathcal{O} (z) \,,  \\[1.0 ex]
\phi(z) &= - {1\over (z-1)^2} + {1\over z-1} \Bigl( {a-2\xi\over 1-\xi}\Bigr)
+ {a \, (\xi-2) + \xi\, ( 4 - 3\xi)\over (\xi-1)^2}
+ \mathcal{O} (z-1) \,, \\[1.0 ex]
\phi(z) &= - {1\over (z-\xi)^2} + {1\over z-\xi} \Bigl( {a-2\over
\xi(\xi-1)}\Bigr) + {a-3 + 4\xi - 2a\, \xi\over \xi^2 (\xi-1)^2}
+ \mathcal{O} (z-\xi) \,, \\[1.0 ex]
\phi(t) &=  -{1\over t^2} + {1\over t} \bigl( a -2 -2\xi) +
(a-3-2\xi+ a\, \xi - 3\xi^2)
+\mathcal{O} (t) \,,
\end{split}
\end{equation}
where $t=1/z$ is used to describe the fourth puncture.  Given a
Strebel quadratic differential
$\varphi = \phi(z) (dz)^2$
that near $z_0$ looks like
\begin{equation}
\phi(z)  = -{1\over (z-z_0)^2} + {r_{-1}\over (z-z_0)} +  r_0 + r_1
(z-z_0) + \ldots\,,
\end{equation}
a canonical local coordinate $w$ (defined up to a phase) is obtained
by requiring
$\varphi = -(1/w^2) (dw)^2$.  This gives~\cite{Moeller:2004yy}:
\begin{equation}
\label{moellerce}
z = z_0 + \rho \, w + {1\over 2} \, r_{-1} (\rho w)^2  + {1\over 16}(7
r_{-1}^2 + 4 r_0) (\rho w)^3 +
\ldots \,,
\end{equation}
where $\rho$ is the mapping radius, which can also be obtained using
the quadratic
differential.
Comparing (\ref{coordexp}) and (\ref{moellerce}) we see that $\beta = r_{-1}/2$
at each punture.
We can therefore use the expansions (\ref{thefirstorderpole})
to read:
\begin{equation}
\label{betavalues}
\beta_1={a\over 2\xi} -{1\over \xi} -1 \,,\quad
\beta_2=\frac{a-2\xi}{2(1-\xi)}\,,\quad
\beta_3=\frac{a-2}{2\xi(\xi-1)}\,,\quad
\beta_4= {a\over 2} -1 -\xi \,.
\end{equation}
Since $a$ is a function of $\xi$ and $\bar \xi$, all the $\beta_I$ are
functions of $\xi$ and $\bar \xi$.

We can now proceed to get the values of the coordinate
expansion coefficients $\gamma$ in terms of $a$ and $\xi$.
As noted in (\ref{coeffs1}), however, the quantity
$\varepsilon= 8\beta^2-6\gamma$ is more useful. A short
calculation shows that $\varepsilon = -(5 \beta^2 + 3r_0)/2$.
Reading the various values of $r_0$ from the expansions
(\ref{thefirstorderpole}) and the various values of $\beta$ from
(\ref{betavalues}) we find
\begin{equation}
\label{epsvalues}
\begin{split}
\varepsilon_1&= 2+{1\over \xi} (a-2) + {1\over \xi^2}
\Bigl( 2 + a - {5\over 8} a^2\Bigr)\,, \\[0.3ex]
\varepsilon_2&= {-5a^2 + 16\, \xi(\xi -3) + 8a \,(\xi + 3)\over
8\, (\xi-1)^2}\,,
\\[0.3ex]
\varepsilon_3&= {16+ 8a -5a^2 + 24(a-2)\xi \over 8\,\xi^2 \,(\xi-1)^2}\,,
\\[0.3ex]
\varepsilon_4&= 2+ a - {5\over  8} \, a^2 - 2 \xi + a\, \xi + 2 \xi^2 \,.
\end{split}
\end{equation}

The function $a(\xi)$ is known
numerically to high accuracy for $\xi\in \mathcal{A}$, where
$\mathcal{A}$ is a specific subspace of $\mathcal{V}_{0,4}$
described in detail in Figures 3 and 6 of ref.~\cite{Moeller:2004yy}. The
full space $\mathcal{V}_{0,4}$ is obtained by acting on
$\mathcal{A}$ with the transformations generated by
$\xi\to 1-\xi$ and $\xi\to 1/\xi$, together with complex
conjugation $\xi\to
\bar\xi$. In fact $\mathcal{V}_{0,4}$
contains twelve copies of $\mathcal{A}$.  Let $f(\mathcal{A})$
denote the region obtained by mapping each point $\xi\in
\mathcal{A}$ to $f(\xi)$. Then $\mathcal{V}_{0,4}$ is composed of the six
regions
\begin{equation}
\label{various_regions}
\mathcal{A} \,, ~~ {1\over \mathcal{A}}\,,~~ 1-\mathcal{A}\,,
  ~~{1\over 1-\mathcal{A}} \,, ~~1- {1\over \mathcal{A}}\,,~~
{\mathcal{A}\over 1-\mathcal{A}} \,,
\end{equation}
together with their complex conjugates. The transformations
$\xi\to 1-\xi$ and $\xi\to 1/\xi$ are
$SL(2, \mathbb{C})$ transformations that permute the points
$0, 1,$ and $\infty$.  While doing so, they move the third puncture
among the various regions in (\ref{various_regions}).
The assignment of coordinates to punctures must be consistent
with the $SL(2, \mathbb{C})$ transformations that exchange the
punctures: the quadratic differential
on two conformally related surfaces must agree.
For example,
letting $\tilde z= 1-z$, we can calculate $\phi(\tilde z)$ using
$\phi(\tilde z) d\tilde z^2 = \phi(z) dz^2$.  We must find that
$\phi(\tilde z)$ takes the form in (\ref{strebelqd}) with $\xi$ replaced
by $1-\xi$, and $a(\xi)$ replaced by $a(1-\xi)$.  Completely analogous
remarks hold for the transformation $\tilde z = 1/z$.  Doing these
transformations explicitly we find
\begin{equation}
a\,(1-\xi)= 4-a(\xi)\,,\quad
a\Bigl(\,{1\over\xi}\,\Bigr)= {a(\xi)\over \xi}\,.
\end{equation}
These equations define $a$ over the full set of regions
in (\ref{various_regions}) once it is given on $\mathcal{A}$.

The reality of the string field theory action is guaranteed
if the local coordinates on surfaces that are mirror images
of each other
are related by the (antiholomorphic) mirror map.
Consider two four-punctured spheres: the first with uniformizer $z$ and third
puncture at  $\xi$,
the second with uniformizer $\tilde z$ and third puncture at  $\bar\xi$.
The antiholomorphic map relating the punctured spheres is $\tilde z =
\bar z$. Two local
coordinates $w$ and $\tilde w$ are mirror related if $\tilde w (p^*)
= \overline{w(p)}$,
where $p$ is a point and $p^*$ is its image under the mirror map.
In order to
obtain mirror-related local coordinates the
associated quadratic differentials on the surfaces must satisfy
$\phi (\tilde z) =  \overline{\phi (z)}$~\cite{strebel}.  It follows
from (\ref{strebelqd}) that $\phi(\tilde z)$ takes the form indicated
in this equation, with $z$ replaced by $\tilde z$ and   $a(\xi)$
replaced by $\overline{a(\xi)}$.  We thus learn that
\begin{equation}
a(\,\bar\xi\,)=\overline{a(\xi)} \,.
\end{equation}
This definition guarantees that the contribution to the amplitude
of any region $\mathcal{S}\subset\mathcal{V}_{0,4}$ and the contribution
from $\overline{\mathcal{S}}$ are complex conjugates of each other.
Consequently, the integral over $\mathcal{V}_{0,4}$ can be done
by integrating over the six regions in (\ref{various_regions}) and
adding to the result its complex conjugate.

With $a(\xi)$ now defined over $\mathcal{V}_{0,4}$, we can also
find formulae that define the mapping radii $\rho_I$ and the
coordinate expansion coefficients $\beta_I$ on the various copies
of $\mathcal{A}$ in terms of the values on $\mathcal{A}$.  These
formulae are given in Appendix B.

\subsection{Quartic interactions for states with $c\bar c$ ghost factor}

In order to illustrate the earlier discussion we
consider an important class of relatively simple
four-point amplitudes.
Suppose we want to evaluate the amplitude $\{M_1, M_2, M_3,
M_4\}$ with states $M_i$ of the form
\begin{equation}
|M_i\rangle  = \mathcal{O}_i  c_1 \bar c_1 |0\rangle,
\end{equation}
where $\mathcal{O}_i$ is some expression built with
matter oscillators.  One can
see that the ghost part of $M_i$ is the same as that of
the tachyon field. First consider the antighost insertion
$\mathcal{B}\, \mathcal{B}^\star$.  Since all the states have
ghost oscillators with mode number one, only the $b_{-1}, \bar b_{-1}$
part of the antighost insertion is relevant.  Using (\ref{Bgeneral})
we see that:
\begin{equation}
\mathcal{B}\, \mathcal{B}^\star =  B^3_{-1} b^{(3)}_{-1}  \, \,
\overline{B^3_{-1}}\, {\bar b}^{(3)}_{-1} + \cdots
\end{equation}
where the dots refer to terms that are not needed in this calculation.
It follows that
\begin{eqnarray}
\mathcal{B}\, \mathcal{B}^\star (c_1 \bar c_1)^{(1)} (c_1 \bar c_1)^{(2)}
(c_{1} \bar c_{1})^{(3)}
(c_1 \bar c_1)^{(4)} |0\rangle &=&
-B_{-1}^3\overline{B_{-1}^3} (c_1 \bar c_1)^{(1)} (c_1 \bar
c_1)^{(2)} (c_1 \bar c_1)^{(4)}|0\rangle. \nonumber\\[1.0ex]
&=&-\frac{1}{\rho_3^2}(c_1 \bar c_1)^{(1)} (c_1 \bar c_1)^{(2)}
(c_1 \bar c_1)^{(4)}|0\rangle.
\end{eqnarray}
Note that in our convention the states with superscripts
$1,2,3$, and $4$ are inserted at $z=0,1,\xi$, and $\infty$,
respectively. The ghost part of the overlap is then
\begin{equation}
\label{trivialst} \langle \Sigma|\mathcal{B}\,
\mathcal{B}^\star (c_1 \bar c_1)^{(1)} (c_1 \bar c_1)^{(2)} (c_{1}
\bar c_{1})^{(3)} (c_1 \bar c_1)^{(4)}
|0\rangle
=-\frac{1}{(\rho_1\rho_2\rho_3\rho_4)^2}\langle c \bar
c(z_1)c \bar c(z_2)c \bar c(t=0)\rangle\,,
\end{equation}
where the conformal transformation of
each ghost oscillator introduces a factor of the mapping radius.
For the fourth puncture, which is located at $t=1/z=0$,
the mapping radius $\rho_4$ refers to the $t$ coordinate
and it is a finite number.
To compute the above correlator we note that
\begin{equation}
c\bar c (t=0) = \lim_{z\to \infty} {1\over |z|^4} \, c \bar c (z)\,,
\end{equation}
and therefore
\begin{equation}
\langle c \bar
c(z_1)c \bar c(z_2)c \bar c(t=0)\rangle =
\lim_{z\to \infty} {1\over |z|^4} \langle c \bar
c(z_1)c \bar c(z_2)c \bar c(z)\rangle
= 2\lim_{z\to \infty} {|z_{12}(z_1-z) (z_2-z)|^2 \over |z^4|} =2\,,
\end{equation}
once we set $z_1=0$ and $z_2=1$.  Back in (\ref{trivialst}) we find
\begin{equation}
\label{trivighost} \langle \Sigma|\mathcal{B}\,
\mathcal{B}^\star (c_1 \bar c_1)^{(1)} (c_1 \bar c_1)^{(2)} (c_{1}
\bar c_{1})^{(3)} (c_1 \bar c_1)^{(4)}
|0\rangle
=-\frac{2}{(\rho_1\rho_2\rho_3\rho_4)^2}\,.
\end{equation}

The matter part of the correlator is computed
using (\ref{coordexp}) to map the operators to
the uniformizer $z$, at which stage the correlator is computable.
On the four-punctured sphere $\Sigma_\xi$ with modulus $\xi$ we write:
\begin{equation}
\langle\langle \mathcal{O}_1\mathcal{O}_2\mathcal{O}_3
\mathcal{O}_4\rangle\rangle_\xi \equiv \langle h_1\circ \mathcal{O}_1
~ h_2\circ
\mathcal{O}_2 ~h_3\circ \mathcal{O}_3 ~h_4\circ \mathcal{O}_4
\rangle_{\Sigma_\xi},
\end{equation}
where the right-hand side is a matter correlator computed after
the local operators $\mathcal{O}_i$ have been mapped.
Our final result is therefore:
\begin{equation}
\{ M_1, M_2, M_3, M_4\} = -{2\over \pi }\int_{\mathcal{V}_{0,4}}
{dx dy\over
(\rho_1\rho_2\rho_3\rho_4)^2}  \,\langle\langle
\mathcal{O}_1\mathcal{O}_2\mathcal{O}_3
\mathcal{O}_4\rangle\rangle_\xi \,.
\end{equation}

For the case of four zero-momentum tachyons $T = c_1 \bar c_1 |0\rangle$
the matter operators are the identity and the
matter correlator is equal to one. We thus get
\begin{eqnarray}
\{T^4\}=-\frac{2}{\pi} \int_{\mathcal{V}_{0,4}} {dx
dy \over (\rho_1\rho_2\rho_3\rho_4)^2}  \,.
\end{eqnarray}
The integrand is manifestly real.  Since four identical states
have been inserted at the punctures, the measure to be integrated
must be fully invariant under the $SL(2, \mathbb{C})$ transformations
that generate the six regions in (\ref{various_regions}). Therefore
the full integral is equal to 12 times the integral over $\mathcal{A}$.
This integral is easily done numerically using the data
given in~\cite{Moeller:2004yy} and we recover the familiar value
\begin{equation}
\label{tachresoefkv}
\{T^4\} = -\frac{24}{\pi} \int_{\mathcal{A}} {dx
dy\over (\rho_1\rho_2\rho_3\rho_4)^2}
=-72.414\, .
\end{equation}
The corresponding term in the string field potential is
$\kappa^2 V=\frac{1}{4!}\{T^4\}\,t^4=-3.0172\, t^4.$

\sectiono{The explicit computation of quartic couplings}\label{sect4}

  The
terms $a^4$, $a^2 d^2$, and $d^4$ in the string field potential
receive contributions from cubic interactions
of all levels.  They also receive
contributions  from
the elementary quartic vertex.
Since $a$ and $d$ are marginal directions, these
two types of contributions must cancel. In our earlier
paper~\cite{Yang:2005iu} we verified
that this cancellation holds with good accuracy for the
$a^4$ term.    We noted that
a potential complication:  in closed string field
theory the cubic vertex does not fully
determine the quartic vertex, so the cancellation should happen
for all four-string vertices that are consistent with the
cubic vertex. This will happen if the integrands of the
four-point amplitudes are total derivatives.
In this case the quartic amplitude arises from
the boundary of $\mathcal{V}_{0,4}$. This boundary is completely
determined by the geometry of the three-string vertex: gauge
invariance requires  that the boundary of $\mathcal{V}_{0,4}$
match the configurations obtained with two cubic vertices joined
by a collapsed propagator.
Letting $G$ and $D$ denote the states associated with $a$ and $d$
respectively (see (\ref{thefirstexp})), the integrands
in $\{G^4\}$, $\{G^2 D^2\}$, and $\{D^4\}$ are thus expected
to be total derivatives. In our earlier paper we confirmed that the
integrand in $\{G^4\}$ is a total derivative.
We will do the same here for the other two amplitudes.

The computations to be discussed below determine the
quartic contribution to the potential.  The results we obtained
are summarized by
\begin{equation}
\label{v4value}
\kappa^2 V_4 (a,d) =  -0.2560\, a^4  + 0.4571 \, a^2 d^2  -0.1056 \, d^4 \,.
\end{equation}

\subsection{Elementary contribution to $a^2 d^2$}

In order to compute this amplitude we insert $G$
at the first and fourth punctures ($z=0$ and $z=\infty$, respectively)
and $D$ at the second and third punctures ($z=1$ and $z=\xi$,
respectively). We begin our analysis with the computation of the
ghost part of the amplitude.

Consider the antighost insertion $\mathcal{B}\mathcal{B}^\star$ acting
on the ghost part of the four states:
\begin{equation}
\label{theghosttdd}
\mathcal{B}\, \mathcal{B}^\star (c_1\bar c_1)^{(1)}
(c_1 c_{-1} -\bar c_1\bar c_{-1})^{(2)} (c_1 c_{-1}
-\bar c_1\bar
c_{-1})^{(3)}(c_1\bar
c_1)^{(4)}|0\rangle \,,
\end{equation}
Since punctures one and four are fixed and the states inserted
in them have ghost oscillators $c_1 \bar c_1$, the antighost factor
$\mathcal{B}\mathcal{B}^\star$ is only supported on punctures two
and three.  A small calculation shows that
\begin{equation}
\mathcal{B}\, \mathcal{B}^\star
(c_1 c_{-1} -\bar c_1\bar c_{-1})^{(2)} (c_1 c_{-1}
-\bar c_1\bar
c_{-1})^{(3)}=({\Ci_1^{2}}\overline{\Ci_1^3}-B_1^{2}
\overline{B_1^{3}})c_1^{(2)} \bar c_1^{(3)}+ B_{1}^{2}
\overline{B^3_{-1}}\, c_1^{(2)}\bar c_{-1}^{(3)}\, +\,\star\hbox{-conj}\,.
\end{equation}
Both sides of this equation have vacuum states to the right, which have
not been written to avoid clutter.  It follows that
\begin{equation}
\label{organizeamp}
\begin{split}
\langle\Sigma|\mathcal{B}\mathcal{B}^\star |T\rangle |D\rangle |D\rangle
|T\rangle &= ({\Ci_1^{2}}\overline{\Ci_1^3}-B_1^{2}
\overline{B_1^{3}})\,
\langle (c_1\bar c_1)^{(1)} \,, c_1^{(2)} \,,  \bar c_1^{(3)}\,,
(c_1\bar c_1)^{(4)}\rangle
\\[1.0 ex]
&~~+B_{1}^{2}
\overline{B^3_{-1}}\,
\langle (c_1\bar c_1)^{(1)} \,, c_1^{(2)} \,,  \bar c_{-1}^{(3)}\,,
(c_1\bar c_1)^{(4)}\rangle
\,+\, *\hbox{-conj}\,.
\end{split}
\end{equation}
Note that the star-conjugate insertions give rise to complex
conjugate correlators. This happens because all other ghost states
in the correlator are self-conjugate.  Two correlators are thus needed to
evaluate the (\ref{organizeamp}). The first arises from the first line
on the right-hand side
\begin{equation}
\label{firstcoore}
\langle (c_1\bar c_1)^{(1)} \,, c_1^{(2)} \,,  \bar c_1^{(3)}\,,
(c_1\bar c_1)^{(4)}\rangle
= {2\bar\xi\over \rho_1^2 \rho_2\rho_3\rho_4^2} \,.
\end{equation}
The second correlator appears on the second line
of (\ref{organizeamp}). It involves the state created by $\bar
c_{-1}^{(3)}$ on the vacuum. The corresponding operator ${1\over
2} \bar\partial^2 \bar c$ has a nontrivial transformation under a
conformal map:
\begin{equation}
{1\over 2} \bar\partial^2 \bar c (\bar w) = \rho_I \Bigl(
{1\over 2} \bar\partial^2 \bar c (\bar z_I) -\bar\beta_I \bar\partial
\bar c (\bar z_I)
+ {\bar\varepsilon_I\over 2} \bar c (\bar z_I)\Bigr) \,.
\end{equation}
With the help of this transformation we find
\begin{equation}
\label{secibdcoore}
\langle (c_1\bar c_1)^{(1)} \,, c_1^{(2)} \,,  \bar c_{-1}^{(3)}\,,
(c_1\bar c_1)^{(4)}\rangle
= {\rho_3\over \rho_1^2 \rho_2\rho_4^2}\, ( \bar \varepsilon_3 \bar
\xi - 2 \bar \beta_3) \,.
\end{equation}
Finally, we simplify the expressions in (\ref{organizeamp}) that
depend on the coefficients $B$ and $C$:
\begin{equation}
{\Ci_1^{2}}\overline{\Ci_1^{3}}-{B_1^{2}}\overline{B_1^{3}} =\rho_2
\rho_3\Big(\bar\p \beta_2\p\bar\beta_3 -\p\beta_2 \bar\p
\bar\beta_3-\frac{1 }{2}\bar\varepsilon_3 \p \beta_2\Big)\,, \qquad
{B_1^{2}}  \overline{B_{-1}^{3}} =\frac{\rho_2}{\rho_3}\,\p
\beta_2\,.
\end{equation}
Using the above results (\ref{organizeamp}) simplifies down to
\begin{equation}
\label{ghpart2dil}
\langle\Sigma|\mathcal{B}\mathcal{B}^\star |T\rangle |D\rangle |D\rangle
|T\rangle = {2\over (\rho_1\rho_4)^2} \Bigl(  \bar\p \beta_2
\p(\bar\xi \bar \beta_3) -
\p\beta_2 \bar\p ( \bar \xi \bar \beta_3) + *\hbox{-conj}\Bigr)\,.
\end{equation}
This concludes our computation of the ghost part of the integrand.

The matter part of the integrand is much simpler.  We have two $G$'s
one at $z=0$ and one at $z=\infty$.  A short computation gives
\begin{equation}
\langle\langle (\p X\bar\p X)^{(1)}(\p X\bar\p
X)^{(4)}\rangle\rangle=(\rho_1\rho_4)^2 \,.
\end{equation}
Note that  the
powers of mapping radii cancel out
  in the product of ghost and matter amplitudes.
Making use of (\ref{amplitude}) the four-point amplitude is
\begin{eqnarray}
\{G^2D^2\}=\frac{4}{\pi}\int_{\mathcal{V}_{0,4}} dx  dy ~
\hbox{Re}\Big(\bar\p \beta_2 \p(\bar\xi \bar \beta_3) - \p\beta_2
\bar\p ( \bar \xi \bar \beta_3)\Big)\,.
\end{eqnarray}
Since we have the same states on punctures one and four, and these
punctures are exchanged by the transformation $z\to 1/z$, the
integral over $\mathcal{A}$ gives the same contribution as the
integral over $1/\mathcal{A}$.  This follows from the $SL(2, \mathbb{C})$
invariance of the construction, but  can also be checked explicitly
using the formulae given in Appendix~B. Since the integrand is
already real, conjugate regions give identical contributions.
Consequently, the four regions $\mathcal{A},\, 1/\mathcal{A},
\,\overline{\mathcal{A}},$ and $1/\overline{\mathcal{A}}$ all give
the same contribution.  To get the full amplitude we must multiply
the contributions of $\mathcal{A}$, of $1-\mathcal{A}$ and
$1-1/\mathcal{A}$ by four:
Therefore, the full amplitude is:
\begin{eqnarray}
\label{builda2d2}
\{G^2D^2\}&=&4\cdot \frac{4}{\pi}\,\Bigl[\,\int_{\mathcal{A}}
+\int_{1-\mathcal{A}}+ \int_{1-1/\mathcal{A}}\,\Bigr]dx  dy ~
\hbox{Re}\Big(\bar\p \beta_2 \p(\bar\xi \bar \beta_3) - \p\beta_2
\bar\p ( \bar \xi \bar \beta_3)\Big)\nonumber\\[1.3ex]
&\simeq & 4\, (-0.03122 + 0.09671 + 0.39157)   =   1.8283.
\end{eqnarray}
In the above, the second and third integrals were evaluated by pulling
back the integrands into $\mathcal{A}$, where all relevant functions
are known numerically.  The details are given in Appendix~B.
The contribution to the potential is
\begin{equation}
\kappa^2 V=\frac{6}{4!}\{G^2D^2\}\,a^2 d^2 =0.4571 \, a^2 d^2.
\end{equation}

\medskip

We now verify that the integrand of the $\{ G^2 D^2\}$
amplitude is a total derivative. Indeed, a short computation
shows that
\begin{equation}
\bigl(\bar\p \beta_2 \p(\bar\xi \bar \beta_3) - \p\beta_2
\bar\p ( \bar \xi \bar \beta_3)\bigr)~d\xi \wedge d\bar\xi  =
d\, \Omega^{(1)}\,,
\end{equation}
where
\begin{equation}
\Omega^{(1)} = {1\over 2} \,\bigl[ -\beta_2 \partial (\bar\xi\bar\beta_3)
+ (\partial \beta_2 )\, \bar\xi \bar \beta_3\,\bigr] \, d\xi
+{1\over 2} \,\bigl[\, (\bar\p \beta_2) \,\bar\xi\bar\beta_3
-\beta_2  \bar\p (\bar\xi \bar \beta_3)\,\bigr] \, d\bar\xi\,.
\end{equation}
Since $a(\xi)$, $\beta_2(\xi),$ and $\beta_3(\xi)$ are all regular
functions on $\mathcal{V}_{0,4}$, we see that $\Omega^{(1)}$ is
a well-defined one-form. The amplitude reduces to the integral
of $\Omega^{(1)}$ over the boundary of $\mathcal{V}_{0,4}$.

\subsection{Elementary contribution to $d^4$}

For the amplitude $\{ D^4\}$ the matter correlator is just one but
the ghost correlator is quite nontrivial. The antighost insertion
$\mathcal{B}\mathcal{B}^\star$ acts on four
dilaton states. Ghost number
conservation implies that the only nonvanishing correlators are those
in which the antighost insertion supplies one $b$ oscillator
and one $\bar b$ oscillator.  Note also that the term
$b_{-1}^{(3)}\bar b_{-1}^{(3)}$ does not contribute.
Making use of (\ref{Bgeneral}) the relevant terms in
the antighost insertion are
\begin{equation}
\mathcal{B}\mathcal{B}^\star = \Bigl(\,B_{-1}^{3}b_{-1}^{(3)}\sum_{I\not= 3}
\overline{B^{I}_1}\, \bar b_1^{(I)}  + \,\star\hbox{-conj.}\Bigr)
+ \sum_{I\not=J} M^{IJ}\, b_1^{(I)}
\bar b_1^{(J)} \,, \quad
M^{IJ} \equiv B_1^I \overline{B_1^J} -C_1^I  \overline{C_1^J} \,.
\end{equation}
Acting on the four dilatons and forming the correlator,
\begin{equation}
\label{4dilamp}
\begin{split}
\langle \Sigma|\mathcal{B}\mathcal{B}^\star |D^4\rangle
  &= \Bigl(
\sum_{I\not= 3} \,B_{-1}^{3} \overline{B^{I}_1}
\,\bigl\langle \, D, D, c_{-1}^{(3)} , \bar c_1^{(I)} \bigr\rangle +
\,\hbox{c.c.}
  \Bigr)
- \sum_{I\not=J} M^{IJ}
\bigl\langle  D, D, c_1^{(I)} \,, \bar c_1^{(J)} \bigr\rangle \\[1.0ex]
&=~ - \hskip-6pt\sum_{I\not=J\not= K \not= 3}
\hskip-6pt B_{-1}^{3} \overline{B^{I}_1}
\,~\bigl\langle \, (c_{-1} c_1)^{(J)}, (\bar c_{-1} \bar c_1)^{(K)},
c_{-1}^{(3)}
, \,\bar c_1^{(I)} \bigr\rangle +
\,\hbox{c.c.} \\[1.0ex]
&\quad ~~+ \hskip-6pt\sum_{I\not=J\not= K \not= L}\hskip-4pt M^{IJ}
~\bigl\langle
(c_{-1} c_1)^{(K)}, (\bar c_{-1} \bar c_1)^{(L)}, c_1^{(I)} \,,
\bar c_1^{(J)} \bigr\rangle  \, .
\end{split}
\end{equation}
In order to complete our calculation we must evaluate the two correlators
that appear on the final right-hand side.  Defining
\begin{equation}
A_{IJ} \equiv \bigl\langle~ (c_{-1} c_1)^{(I)} \,, c_{-1}^{(J)}
\bigr\rangle \,,
\qquad
B_{IJ} \equiv \bigl\langle~ (c_{-1} c_1)^{(I)} \,, c_{1}^{(J)} \bigr\rangle \,,
\end{equation}
the amplitude in (\ref{4dilamp}) becomes
\begin{equation}
\label{44dilamp}
\langle \Sigma|\mathcal{B}\mathcal{B}^\star |D^4\rangle
=~  \hskip-6pt\sum_{I\not=J\not= K \not= 3}
\hskip-6pt 2B_{-1}^{3}\,( \overline{B^{I}_1}
A_{J3} \overline{B_{KI}} +
\,\hbox{c.c.})
- \hskip-6pt\sum_{I\not=J\not= K \not= L}\hskip-4pt 2M^{IJ}
~B_{KI} \overline{B_{LJ}}  \, .
\end{equation}
It just remains to evaluate $A_{IJ}$ and $B_{IJ}$. There is one
small complication here.  Since we treat the fourth puncture
asymmetrically we must distinguish the case when
$I$ or $J$ are equal to four.  We find that for $I,J \not=4$
  \begin{equation}
\label{aijbij}
\begin{split}
A_{IJ} &= \rho_J \Bigl( \beta_J - \beta_I - 2 \beta_I \beta_J z_{IJ}
+ {1\over 2}
\varepsilon_J z_{IJ} ( 1- \beta_I z_{IJ}) \Bigr) \\
B_{IJ} &={1\over \rho_J} \, z_{IJ} \, ( 1- \beta_I \, z_{IJ}) \,,  \qquad
\hbox{for} \quad I,J \not=4\,.
\end{split}
\end{equation}
Here $z_{IJ} = z_I - z_J$. We also need the following special values which
arise when one of the states is located at the fourth puncture ($z=\infty$):
\begin{equation}
\label{special_values}
A_{4J} = \rho_J \Bigl(~ {1\over 2} \,\varepsilon_J ( \beta_4 + z_J) -
\beta_J\Bigr)\,,
\quad  B_{I4} = {\beta_I\over \rho_4}\,, \quad
B_{4J} = {1\over \rho_J} (z_J + \beta_4) \,.
\end{equation}
In these equations $I$ and $J$ are different from four.
It is possible to obtain (\ref{special_values}) by taking a suitable limit
of (\ref{aijbij}).  One must let
\begin{equation}
\label{lim4beta}
\beta_4 \to {1\over z_4} - {\beta_4\over z_4^2}\,, \quad \hbox{and}
\quad  \rho_4 \to z_4^2 \, \rho_4 \,,
\end{equation}
and then take the limit as $z_4 \to \infty$.  The above replacements are
the ones involved in changing uniformizer from $z$ to $t=1/z$.

Having calculated explicitly all the quantities that enter into
(\ref{44dilamp}), we can do the numerical integration:
\begin{equation}
\{D^4\}=\frac{12}{\pi}\int_{\mathcal{A}} dx dy~ \langle
\Sigma|\mathcal{B}\, \mathcal{B}^\star |D^4\rangle
=-2.5336\,.
\end{equation}
The contribution to the potential is therefore
\begin{equation}
\kappa^2 V=\frac{1}{4!}\{D^4\}\,d^4=-0.1056\,  d^4.
\end{equation}

In order to demonstrate that the integrand is a total derivative, we have
shown that
$\langle \Sigma|\mathcal{B}\mathcal{B}^\star |D^4\rangle =
\partial f + \bar\p \bar f $
for a suitable function $f$ of $\xi$ and $\bar \xi$.   As a
result, the two-form integrand (up to overall constants) is
indeed exact:
$d\xi\wedge d\bar \xi~ (\partial f + \bar\p \bar f ) = d \,(  f d\xi - \bar f
d\bar\xi)\,.$
The calculation of $f$ is quite laborious and is best done using
equation (\ref{aijbij}) for all values of $I$ and~$J$.  The result is
then:
\begin{equation}
\label{totderd4}
\begin{split}
f(\xi,\bar\xi)&=-\hskip-6pt\sum_{I\not=J\not=K\not=L}\Bigl\{ \beta_J
\bar\p\bar \beta_I z_{JK}\bar z_{IL}-\beta_J\bar\beta_I
z_{JK}\delta_{3L}
+\beta_J\bar\beta_I z_{JK}\delta_{3I}\\[0.5ex]
&\qquad~~+\frac{1}{2}\beta_J\beta_K\bar\p\bar\beta_I z_{JK}^2\bar z_{IL}
+\frac{1}{2}\beta_J\beta_K \bar\beta_I z_{JK}^2
\delta_{3I}-\frac{1}{2} \beta_J\beta_K \bar\beta_I z_{JK}^2
\delta_{3L}\\[0.5ex]
&\qquad~~+\beta_J\bar \beta_L\bar\p\bar\beta_I z_{JK} \bar
z_{IL}^2+2\beta_J\bar\beta_I\bar\beta_L z_{JK}\bar
z_{IL}\delta_{3I} +\frac{1}{2}\beta_J\beta_K \bar\beta_L
\bar\p\bar\beta_I z_{JK}^2\bar z_{IL}^2\\[0.5ex]
&\qquad~~+ \beta_J\beta_K \bar\beta_I\bar\beta_Lz_{JK}^2\bar
z_{IL}\delta_{3I}- \bar\beta_J\bar\p\beta_I z_{IL}\bar z_{JK}-
\frac{1}{2}\bar\beta_J\bar\beta_K\bar\p\beta_I z_{IL}\bar
z_{JK}^2\\[0.5ex]
&\qquad~~-\beta_L\bar\beta_J\bar\p\beta_I z_{IL}^2 \bar
z_{JK}-\frac{1}{2}\beta_L\bar\beta_J \bar\beta_K \bar\p\beta_I
z_{IL}^2 \bar z_{JK}^2\Bigr\}.
\end{split}
\end{equation}
In this expression all $\beta_4$ must be replaced as indicated in
(\ref{lim4beta}).  The final answer is finite and well defined over
$\mathcal{V}_{0,4}$.

\sectiono{Analysis and conclusions}\label{sect5}

In this final section we discuss our results.  We first examine
the cancellations between cubic and quartic terms, using certain
projections of the cubic data in order to estimate the effects
of terms that have not been computed.  We then attempt to give
a definition of level suitable for quartic interactions.  While
we are not able to give convincing evidence for any specific
definition, we find out that the level suppression observed for
cubic interactions seems to extend to quartic interactions.
This is good news, as it suggests that computations carried out
with cubic and quartic interactions would converge as the level
is increased.

\subsection{Cancellations and fits}

In this work we  checked explicitly
the quartic structure of closed string field theory.
The existence of flat directions implies that the
infinite-level cubic contribution to the effective potential
for $a$ and $d$  must
be cancelled by elementary quartic interactions.
We claim that the potential $\mathbb{V}(a,d)$, defined as
\begin{equation}
\mathbb{V} (a,d)  \equiv \lim_{\ell\to\infty} \mathbb{V}_{(\ell)} (a,d)
+ V_4 (a,d)\,,
\end{equation}
should vanish identically.
Let us see how well we have checked this.  The relevant
data has been collected on Table~\ref{fit_table}.
Even at $\ell=6$ the pattern of cancellations is
quite clear.  For $a^4$ the quartic term cancels 86\%
of the cubic answer. For $a^2d^2$ and $d^4$ the quartic
interactions  cancel about 90\%  of the cubic answers.

\begin{table}[ht]
\begin{center}
{\renewcommand\arraystretch{1.1}
\begin{tabular}{|c|c|c|c|}
            \hline
            level/vertex
            &~~ $\mathcal{C}_{a^4}$~~
            & $\mathcal{C}_{a^2d^2}$
            &~~ $\mathcal{C}_{d^4}$~~  \\[0.2ex]
            \hline
            $\ell=6$,~\hbox{cubic}   & $\phantom{-}0.2978\phantom{^*}$
             & $-0.5065$&  $\phantom{-}0.1163$\\
            \hline
            $\ell=\infty$, projected
            & $\phantom{-}0.2559\phantom{^*}$
            & $-0.4488$ & $\phantom{-}0.1044$ \\
            \hline\hline
            elementary quartic  & $-0.2560\phantom{^*}$
             & $\phantom{-}0.4571$&  $-0.1056$ \\
            \hline
\end{tabular}}
\end{center}
\caption{\small The coefficients of the quartic terms in the
effective potential
for $a$ and $d$.  First row: result from cubic interactions integrating
massive fields up to level six.  Second row:  projected result from cubic
interactions to all levels.  Third row: elementary quartic contributions,
read from (\ref{v4value}).
} \label{fit_table}
\end{table}

Consider now the same cancellations using projections
from the data of cubic computations.  As we explained in
ref.~\cite{Yang:2005iu}, a fit for the coefficient $\mathcal{C}_{a^4}$
using the best available data gives:
\begin{equation}
\label{fac4}
\mathcal{C}_{a^4}(\ell) \simeq 0.25585 +\frac{0.50581}{\ell^2}
+\frac{1.06366}{\ell^3},
\end{equation}
The above  fit was implied by the fit with leading $1/\ell$
of the related open string coefficients.
  As remarked
in~\cite{Yang:2005iu}, the projected value  0.25585 that follows from
(\ref{fac4})
and the elementary quartic value $(-0.25598)$ essentially agree perfectly.

We do not have  a priori arguments that tell what kind of fit
should be used for $\mathcal{C}_{d^4}$. Equation (\ref{fac4})
would suggest a $1/\ell^2$ fit, and this gives a somewhat low
projection (0.09374). We thus attempted a fit to $1/\ell^3$, which
works very well using the data for $\ell=4$ and $\ell=6$ in
Table~\ref{cubic_table}:
\begin{equation}
\label{cd4fitz}
\mathcal{C}_{d^4}(\ell) \simeq 0.1044 +   \,{2.5647\over \ell^3}  \,.
\end{equation}
The projection 0.1044 is cancelled by the elementary quartic term
($-0.1056$) to an accuracy of 1.1\%.  As a
check of the plausibility of (\ref{cd4fitz}) we attempted a fit
of the form $f_0 + f_1/\ell^\gamma$ and adjusted $\gamma$ so that
$f_0$  matches precisely the elementary quartic term.  This gives
$\gamma \sim 3.2$,
which is reasonably close to our guess $\gamma=3$.  As a further check
we use the
data of Table~\ref{cubic_tablexx}.  Since the contribution
for $\ell=8$ is exceptional, we only use the data for $\ell=4,6,$ and $10$.
This time we  get
\begin{equation}
\label{cd4fit} \mathcal{C}_{d^4}(\ell) \simeq 0.1058 +
\,{2.4593\over \ell^3}  \,.
\end{equation}
This projection is exceptionally good, it cancels the elementary
quartic term to an accuracy of 0.2\%.

We have no guidance for $\mathcal{C}_{a^2d^2}$, either. It seems
reasonable to take a level dependence somewhere in between
those of $\mathcal{C}_{a^4}$ and $\mathcal{C}_{d^4}$.  We thus
considered a fit with $\ell^{-5/2}$ finding:
\begin{equation}
\label{cadfit}
\mathcal{C}_{a^2d^2}(\ell) \simeq -0.4488 -   \,{5.0880\over \ell^{5/2}}  \,.
\end{equation}
The projection $(-0.4488)$ is cancelled by the elementary quartic term
($0.4571$) to an accuracy of 1.8\%. We also found that the match is perfect
for a fit with $\ell^{-\gamma}$ with $\gamma\simeq 2.7$.
All in all, we believe that the above results are good evidence that
the elementary quartic amplitudes  of marginal operators have been
computed correctly.

\begin{figure}
\centerline{\hbox{\epsfig{figure=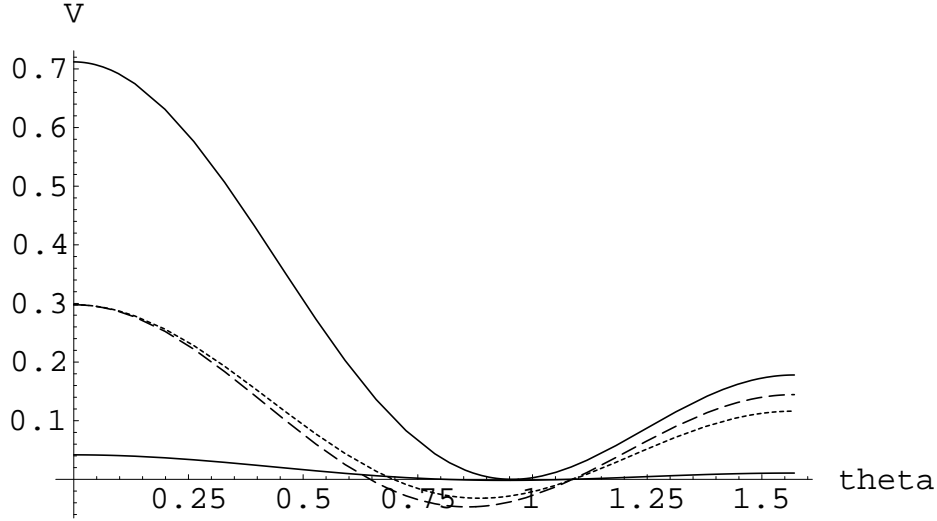,
height=7cm}}} \caption{The potential $\mathbb{V}(a=\cos\theta,d=\sin\theta)$
plotted as a function
of $\theta$.  We show the potentials $\mathbb{V}_{(0)}$ (highest curve
for $\theta=0$), $\mathbb{V}_{(4)}$ (the dashed line),
$\mathbb{V}_{(6)}$ (the dotted line), and the potential including the
quartic elementary interactions.} \label{amassrginal}
\end{figure}

\medskip
The vanishing of the coefficient of $a^2 d^2$ in the effective potential
confirms the expectation that the marginal directions
$a$ and $d$ in fact generate a two dimensional space of marginal directions.
An effective potential with vanishing $a^4$ and $d^4$ terms but nonvanishing
$a^2d^2$ would be consistent with the existence of
marginal directions -- but no two-dimensional moduli space.  It is thus
interesting to visualize the increasing flatness
of the effective potential on the two dimensional space.  Since the
calculated potentials are, to this order, invariant under the
separate tranformations
$a\to -a$ and $d\to -d$, it suffices to consider the potential on the
first quadrant of the $(a,d)$ plane. Since the potential
scales uniformly when $(a,d)\to (\lambda a, \lambda d)$, it suffices
to examine the potential along the arc of a unit circle on the first
quadrant.  Letting $a = \cos\theta$ and $d=\sin \theta$, we can examine
the potentials $\mathbb{V}(a,d)$ as  $\theta \in [0, \pi/2]$.
For this we have produced Figure~\ref{amassrginal}, which plots the potentials
$\mathbb{V}_{(\ell)} (\cos\theta, \sin\theta)$
as a function of $\theta$.  The solid top curve is
$\mathbb{V}_{(0)}(\cos\theta,\sin\theta)$, which happens to
be a perfect square and
vanishes for $\theta = \tan^{-1}(\sqrt{2}) \simeq 0.955$.
While the coefficients $\mathcal{C}_{\dots} (\ell)$ decrease
as we increase $\ell$, the potentials do not approach zero
uniformly as a function of $\theta$:
we do {\em not} have  $|\mathbb{V}_{(0)} (\theta)| \geq
|\mathbb{V}_{(4)} (\theta)| \geq|\mathbb{V}_{(6)} (\theta)|$.
This is especially clear for a range of $\theta$ values near to but smaller
than one.  The final curve we show is $\mathbb{V}_{(6)} + V_4$.  This
curve, solid and near the horizontal axis, makes it clear that even
without projection of the cubic data the effective potential has
become rather small.

The elementary quartic contribution  to $a^4$ (of value $-0.2560$) is
roughly of the magnitude that would be produced by integrating out
fields of level 6.  This is suggested by Table~\ref{cubic_table}, where we see
that
the contribution from fields of level 4 are larger than the quartic
contribution, and contributions from level 8 (of value $-0.0319$) are
considerably smaller (in this rough argument it is not relevant that
the contribution of level six fields is actually zero).   On the
other hand, both for $a^2d^2$ and $d^4$, the quartic contributions
are larger than those that arise from integrating level four fields,
though significantly smaller than those from integrating out the tachyon.
We can safely state that
relevant {\em quartic interactions must be included as the level of the string
field is two or higher}.  For the case at hand, it means including the
quartic interactions after  integrating out the tachyon
in the cubic action.

\subsection{Quartic suppression and attempts to define level}

Here we discuss possible definitions of level for quartic
interactions.  A fully successful definition would have
the following properties:

\begin{enumerate}

\item It should furnish a meaningful comparison between
quartic interactions:  contributions to calculable quantities
from quartic interactions should be suppressed as the level of the interactions
  increases.

\item  It should furnish a meaningful comparison between cubic and
quartic interactions: cubic and quartic
interactions of the same level should have roughly
equal contributions to calculable quantities.

\end{enumerate}

Needless to say, it is easier to satisfy 1 than it is to satisfy 2.
Since it is not understood, even in open
string field theory,  why level expansion works,
we will focus on a property that
is expected to play some role:  amplitudes have level-dependent powers
of mapping radii that give rise to exponential suppression.
This is clearly not the full story in level expansion, where
convergence is typically characterized by corrections with inverse
powers of level~\cite{Taylor:2002fy,Beccaria}. We thus consider the following
a first attempt on a difficult problem.

We begin first with cubic interactions.
We also consider, for simplicity, three
identical states of level $\ell$.
  We will assume that the leading level dependence of the
  three-string coupling is due to the dimension of the state and
the associated conformal map; this gives a factor $\mathcal{R}^{-(L_0
+ \bar L_0)}
= \mathcal{R}^{-\ell + 2}$
for each state. This is the leading level dependence if, (1) the
3-point correlator has at most power dependence on $\ell$
and, (2) the contributions that arise because the operator is not primary
are also suppressed.  It is natural to assume that the level of a cubic
interaction in closed string field theory is defined just like in
open string field theory,  by the sum of levels of the fields that
are coupled. Then, in our present case, the cubic term in the action
has level $L = 3\ell$ and  is of the form
${1\over 3!} c_3 \phi^3$, with
\begin{equation}
\label{c3ui}
c_3 (L) \sim  \mathcal{R}^{-3\ell+ 6} = \mathcal{R}^{-L + 6} = \exp \bigl( -L
\ln \mathcal{R}  + 6 \ln \mathcal{R}\bigr)\,, \quad
\ln \mathcal{R} \simeq 0.2616\,.
\end{equation}
The exponential
suppression due to level is striking.  Even the growth in the number of states
cannot match it:  the number of states grows like
$\exp(a_0\sqrt{L})$, where $a_0$ is a finite positive constant. In terms of
$\ell$, the above reads
\begin{equation}
\label{c3fit}
c_3  \sim   \exp \bigl(
  - 0.7849 \,\ell
  + 1.5697\bigr)\,.
\end{equation}
The $\ell$-independent constant in the exponent should {\em not} be trusted
since our assumptions ignored all powers of $\ell$ and all constants.
If included, $c_3$ takes the correct value for three (level zero) tachyons.
So really, we have
\begin{equation}
\label{39kjdkhj}
c_3 \sim \exp ( -0.2616 L)  \,.
\end{equation}

Let's now consider the elementary four-point interaction of four
identical level $l$ states. These
would go like ${1\over 4!} c_4 \phi^4$ where, under assumptions similar
to those stated before, the value of $c_4$ is roughly
\begin{equation}
c_4  \simeq  {24\over \pi} \int_{\mathcal{A}} dx dy \, (\rho_1\rho_2
\rho_3\rho_4)^{\ell -2} \,.
\end{equation}
The above formula is the obvious generalization of the tachyon
quartic amplitude (\ref{tachresoefkv}),
with which it agrees when $\ell=0$.
We have computed numerically $c_4$ for various values of $\ell \geq 0$ and,
interestingly,
the results are well fit by a decaying exponential:
\begin{equation}\label{c4fit}
c_4 \simeq  \exp \bigl(  -1.135 \,\ell + 4.27\, \bigr)  \,.
\end{equation}
Again, the constant term in the exponential is not reliable and
is only included in order to give the correct answer for the coupling
of tachyons. We have thus learned that
\begin{equation}\label{rec4fit}
c_4 \simeq  \exp \bigl(  -1.135 \,\ell  \bigr)  \,.
\end{equation}
The level $L$ of a four-string interaction increases
with $\ell$ so this result suggests that quartic interactions are suppressed
as the level is increased, the statement of the first condition given
at the beginning of this subsection.  This is grounds for optimism.

How should we define the level $L$ of a four string elementary interaction ?
One natural option would be to take
\begin{equation}
\label{levattempt}
L =  \alpha + \beta \sum_{i=1}^4 \ell_i \,,
\end{equation}
with $\alpha$ and $\beta$ constants to be determined.  If we take $\alpha=0$
and $\beta=1$, the simplest generalization of level for  cubic
interactions, we get
$L= 4\ell$ and (\ref{rec4fit}) becomes $c_4 \sim \exp (-0.2838 L)$,
which is intriguingly similar to (\ref{39kjdkhj}). Thus, for $L=4\ell$ we
get $c_3(L) \sim c_4(L)$.

It is not clear, however, that similar levels should lead to $c_3 \sim c_4$.
Effective potentials, for example, suggest that
$c_3^2$ and $c_4$ give similar contributions to observables.
So, in the spirit of condition 2 we can require that for
similar levels we get similar contributions:
\begin{equation}
\label{theconstlrk}
(c_3 (L))^2 \sim c_4 (L) \,.
\end{equation}
Focusing on level dependent terms and using (\ref{39kjdkhj}), (\ref{rec4fit}),
and $L\sim 4\beta \ell$ we find $\beta \simeq 0.54$.  This suggests
$L \sim {1\over 2} \sum_i \ell_i$ for quartic interactions.  We do not have
sufficient data to test the validity of such relation.
One may attempt to find the value of $\alpha$ in (\ref{levattempt})
  but that requires a
control  over level independent terms in our expansions
that we do not have. Using (\ref{c3fit}) and (\ref{c4fit}) at face
value would give
$\alpha \simeq -2.2$. Ideally we would wish $\alpha \sim 2$, which would
add level to quartic interactions.  The negative $\alpha$ we find
is just a reflection of the fact that the quartic tachyon amplitude is
surprisingly big. In fact, it is so big that it eliminates the critical point
in the potential calculated with quadratic
and cubic terms~\cite{Belopolsky:1994bj}.  Our computations with marginal
directions have suggested a much more benign behavior, one in which quartic
contributions are suppressed with respect to the leading contributions from the
cubic term.

We think that the outlook for level expansion in closed
string field theory is positive.  The above estimates
suggest  that
the same reasons that make higher level cubic interactions
suppressed also make higher level quartic interactions
suppressed. The following strategy would then seem safe:
calculate with cubic interactions to high level until
convergence is clear and a result $A_3$ is obtained.
Then add quartic interactions
increasing the level until convergence occurs again, this time
obtaining a corrected result $A_4$.
Continue in this way with quintic
and higher order interactions
to obtain quantities $A_5, A_6, \ldots$. Throughout the process
we want $A_{i+1} \sim A_i$.  The final result is the limit
of $A_n$ as $n\to \infty$.
In carrying out these calculations one would hope that at each
time one begins a computation including terms of one order higher,
there is a set of low-level interactions at that new order such that
the result obtained including them does not differ greatly from
the result obtained without them.

We have seen that  in the present calculations it makes sense
to add quartic interactions once the string field includes states
of level two or higher.  This strategy  helps  produce
clearer convergence. We do not know, in general,
at what point quartic interactions should be added. Since our present
analysis suggests that convergence will occur anyway, such determination
may not be crucial.  When the computation of cubic interactions
is inexpensive, we may compute a large number of them before including
quartic terms.

\subsection{Open questions}

There are several questions that we have not addressed.
We have not attempted to discuss {\em large} marginal deformations
nor {\em large} dilaton deformations.
In level-expanded open string field theory the Wilson
line deformation parameter encounters
an obstruction for a finite value~\cite{Sen:2000hx}.
In closed string field
theory we have seen (sect.~\ref{dcdsdlv}) that after integrating out the cubic
  tachyon interactions
the radius deformation parameter $a$  has a finite
range, while the dilaton deformation has an infinite range.
Finite ranges appear when the solution (marginal) branch meets
another branch of the equations of motion.
  Since higher level and higher order interactions imply equations of motion
of higher order, the results obtained with the
lowest level cubic interactions may be  modified. Or perhaps not.
We have computed the effects of integrating out
the cubic couplings for the lowest level {\em massive} closed string fields.
For ranges in which the computations are reliable we found no evidence
of a limiting value for the dilaton.  This and further results will
be published in~\cite{yang-zwiebach}.

If closed
string field theory were able to define
dilaton deformations that correspond to infinite
string coupling  it would be a very
exciting result.  For superstrings it would imply
that a (yet to be constructed)  type IIA quantum closed
superstring field theory would contain M-theory in its configuration space.
Most likely infinite string coupling would correspond to $d=\infty$.
If infinite coupling could be reached for some finite value $d=d_0$
the situation
might be even better: M-theory would be obtained as a regular expansion around
a point clearly inside the configuration space of the string
field theory.\footnote{
We thank W.~Taylor for suggesting this possibility.}

Our increased computational ability, due largely to
the results of Moeller,  and the experience gained
in this paper make it interesting to reconsider  the
bulk closed string tachyon potential~\cite{yang-zwiebach}.
We are now able to compute
fairly efficiently any set of quartic
interactions.  While the effective descriptions
based on conformal field theory
indicate that the dilaton potential is just a multiplicative
factor for the bulk tachyon potential, this is not the case
in string field theory.  The string field dilaton and the sigma model dilaton
are related by field redefinitions that involve the tachyon and other
massive fields.  An investigation of the bulk tachyon potential in
string field theory requires the inclusion of the dilaton and the computation
of its off-shell couplings to other massive fields.
Only now we have the technology to do this.

\bigskip
\noindent
{\bf Acknowledgements}

We are grateful to M.~Headrick, Y.~Okawa, N.~Moeller, M.~ Schnabl,
A.~Sen, and W.~Taylor
for their comments and suggestions.

\appendix

\sectiono{Cubic potentials} \label{a1}

The potential for the level-four string field in (\ref{lev4sfml}) is
\begin{equation}
\begin{split}
\kappa^2 V_{(4)}&=
(f_1)^2 + f_1\Bigl( \frac{121}{432} a^2
-\frac{1}{96} d^2 \Bigr)
+\frac{625}{4}(f_2)^2
          +f_2\Bigl( \frac{15625}{1728} a^2
-\frac{15625}{3456} d^2 \Bigr)
\\[1ex]
&-\frac{25}{2}\big[ (f_3)^2 +(
f_4)^2\big] -\big(f_3+ f_4\big) \Bigl( \frac{1375}{864} a^2
-\frac{125}{576} d^2
\Bigr)\\[1ex]
& +4 (r_1)^2 + r_1\Bigl( \frac{27}{16} a^2
-\frac{25}{864} d^2 \Bigr)
-2\big[(r_2)^2+( r_3)^2\big]
+\big(r_2+r_3\big)\Bigl( \frac{11}{16}a^2 +\frac{5}{288}d^2
\Bigr)
  \\[1ex]
& + 25\big[(r_4)^2+(r_5)^2\big]
-\big(r_4+r_5\big)\Bigl(
\frac{125}{32}a^2 +\frac{625}{1728}d^2\Bigr)
+ 2\,r_6r_7-\frac{8}{27}(r_7-r_6) a d
\,.
\end{split}
\end{equation}

\noindent
The level six string field needed for the computation of quartic terms in
the potential for $a$ and $d$~is
\begin{equation}
\begin{split}
|\Psi_6\rangle &=\phantom{+}h_1 \, c_{-2}\bar
c_{-2} +\,h_2  \, \alpha_{-2} \alpha_{-1} \,
\bar\alpha_{-2} \bar\alpha_{-1}  c_1 \bar
c_1
+\,h_3\,
b_{-2} c_{-1}c_1\,\,\bar b_{-2} \bar c_{-1}\bar c_1\nonumber\\
&\quad +(\,g_1\, \,\alpha_{-2}\alpha_{-1} \, c_1 \bar c_{-2} + \tilde
g_1\,
\bar\alpha_{-2}\bar\alpha_{-1}  \, c_{-2} \bar c_1 )
+\, (g_2 \,\, b_{-2} c_{-1} c_1 \,\, \bar c_{-2} \,+ \,\tilde g_2
\,
c_{-2}\,\, \bar b_{-2} \bar c_{-1} \bar c_1 )   \nonumber\\
&\quad +\,g_3 \,\alpha_{-2} \alpha_{-1} \, c_1\,\, \bar b_{-2} \bar
c_{-1} \bar c_1 \,+ \,\tilde g_3\, \bar\alpha_{-2} \bar\alpha_{-1}
\,\,
b_{-2}  c_{-1}  c_1  \,\, \bar c_1\nonumber\\[1.0ex]
&\quad +( g_4\,\alpha_{-2}\bar \alpha_{-2} \,\,c_{-1} c_1\,+\,
        \tilde g_4 \,\alpha_{-2}\bar \alpha_{-2} \,\,  \bar c_{-1}\bar
c_1)
+ (g_5\,\alpha_{-1}\bar \alpha_{-2} \,\,c_{-2} c_1\,+\,
        \tilde g_5 \,\alpha_{-2}\bar \alpha_{-1} \,\,  \bar c_{-2}\bar
c_1)\\[0.3ex]
&\quad + g_6\,\alpha_{-2}\bar \alpha_{-1} \,\,c_{-1} c_1\,\, \bar
b_{-2} \bar c_1 \,+\,
        \tilde g_6 \,\alpha_{-1}\bar \alpha_{-2} \,\,  b_{-2} c_1\,\, \bar
c_{-1} \bar c_1
\nonumber\\[0.3ex]
&\quad +\,g_7\, \alpha_{-1}\bar \alpha_{-1} \,\,c_{-2} c_1\,\, \bar
b_{-2} \bar c_1 \, + \tilde g_7\, \alpha_{-1}\bar \alpha_{-1}\,\,
b_{-2} c_1\,\,\bar c_{-2} \bar c_1
\nonumber\\[0.5ex]
&\quad +\,g_8\, \, b_{-2} \,\,  \bar c_{-2} \bar c_{-1}\bar c_1\, +\,
\tilde g_8\,\, c_{-2} c_{-1}c_1\, \, \bar b_{-2} \,.
\end{split}
\end{equation}
This string field contains states
  of ghost numbers $(-1, 3)$ and $(3, -1)$.
The associated potential~is
\begin{equation}
\begin{split}
\kappa^2 V_{(6)}&=4 h_1 h_3+4 g_2\tilde g_2+16 g_4\tilde g_4+8
g_5\tilde g_6+8g_6\tilde g_5+4 g_7\tilde g_7+4 g_8\tilde g_8
\\[0.5 ex]
& + \frac{50}{729} h_1 d^2+\frac{200}{729} h_3 d^2
+\frac{100}{729} (g_2+\tilde g_2) d^2-\frac{200}{729} (g_8+\tilde
g_8) d^2\\[0.5 ex]
  &+\frac{128}{729} (g_4-\tilde g_4) a d-\frac{160}{729}
(g_5-\tilde g_5) a d-\frac{320}{729} (g_6-\tilde g_6) a d
+\frac{400}{729} (g_7-\tilde g_7) a d\,.
\end{split}
\end{equation}

\sectiono{Transformation laws} \label{a2}

We record the following transformation laws:
\begin{align}
\beta_1 (1-\xi) &= -\beta_2 (\xi) \,,     &\rho_1(1-\xi) &= \rho_2(\xi)\,, \\
\beta_2 (1-\xi) &= -\beta_1 (\xi) \,,    &\rho_2(1-\xi) &= \rho_1(\xi)\,,  \\
\beta_3 (1-\xi) &= - \beta_3 (\xi) \,,   &\rho_3(1-\xi) &= \rho_3(\xi)\,,  \\
\beta_4 (1-\xi) &= - \beta_4 (\xi)- 1 \,,   &\rho_4(1-\xi) &= \rho_4(\xi)\,.
\end{align}

\begin{align}
\beta_1 (1/\xi) &= \beta_4 (\xi) \,,     &\rho_1(1/\xi) &= \rho_4(\xi)\,,  \\
\beta_2 (1/\xi) &= 1-\beta_2 (\xi) \,,    &\rho_2(1/\xi) &= \rho_2(\xi) \,, \\
\beta_3 (1/\xi) &= \xi(1-\xi \beta_3 (\xi)) \,,   &\rho_3(1/\xi) &=
\rho_3(\xi)/|\xi|^2 \,, \\
\beta_4 (1/\xi) &= \beta_1 (\xi) \,,   &\rho_4(1/\xi) &= \rho_1(\xi)\,.
\end{align}

\begin{align}
\beta_1 (1-1/\xi) &= \beta_2 (\xi)-1 \,,     &\rho_1(1-1/\xi) &=
\rho_2(\xi)\,,  \\
\beta_2 (1-1/\xi) &= -\beta_4 (\xi) \,,    &\rho_2(1-1/\xi) &=
\rho_4(\xi) \,, \\
\beta_3 (1-1/\xi) &= \xi(\xi\beta_3(\xi)-1)\,,   &\rho_3(1-1/\xi) &=
\rho_3(\xi)/|\xi|^2 \,, \\
\beta_4 (1-1/\xi) &= - \beta_1 (\xi)- 1 \,,   &\rho_4(1-1/\xi) &=
\rho_1(\xi)\,.
\end{align}

For the benefit of the interested reader we discuss the transformations
invoved in computing the second and third integrals in (\ref{builda2d2}).
Using the variable of integration $\xi' = 1-\xi$ over the region
$1-\mathcal{A}$,
the second integral involves
\begin{equation}
I_2 = \int_{1-\mathcal{A}}dx'  dy' ~
\Big(\bar\p' \beta_2(\xi') \p'(\bar\xi' \bar \beta_3(\xi')) - \p'\beta_2(\xi')
\bar\p' ( \bar \xi' \bar \beta_3(\xi'))\Big)\,.
\end{equation}
Using the transformation rules
\begin{equation}
dx' dy' = dx dy,\quad \p' = -\p,\quad  \bar\p' = - \bar \p,\quad
\beta_2(\xi') = -\beta_1(\xi), \quad \beta_3(\xi') = -\beta_3(\xi)\,,
\end{equation}
we find that $I_2$ can be written as the following integral over $\mathcal{A}$:
\begin{equation}
I_2 = \int_{\mathcal{A}}dx  dy ~
\Big(\bar\p \beta_1 \p((1-\bar\xi) \bar \beta_3) - \p\beta_1
\bar\p ((1- \bar \xi) \bar \beta_3)\Big)\,.
\end{equation}
In this integral the argument of all $\beta$'s is $\xi$.
Using the variable
of integration $\xi'= 1-1/\xi$ over $1-{1/\mathcal{A}}$, the
third integral in (\ref{builda2d2}) is written as
\begin{equation}
I_3 =\int_{1-{1\over \mathcal{A}}}dx'  dy' ~
\Big(\bar\p' \beta_2(\xi') \p'(\bar\xi' \bar \beta_3(\xi')) - \p'\beta_2(\xi')
\bar\p' ( \bar \xi' \bar \beta_3(\xi'))\Big)\,.
\end{equation}
This time we use the following transformation rules
\begin{equation}
dx' dy' = {dx dy\over |\xi|^4} ,\quad \p' = \xi^2\p,\quad  \bar\p' = \bar \xi^2
  \bar \p,\quad
\beta_2(\xi') = -\beta_4(\xi), \quad \beta_3(\xi) =
\xi( \xi\beta_3(\xi)-1) \,,
\end{equation}
and the integral $I_3$ can now be written as an integral over $\mathcal{A}$:
\begin{equation}
I_3 =\int_{\mathcal{A}}dx  dy ~
\Big(-\bar\p \beta_4 \p [(\bar\xi-1) (\bar\xi\bar \beta_3 -1)] + \p\beta_4
\bar\p [(\bar \xi-1) (\bar \xi\bar \beta_3-1)]\Big)\,.
\end{equation}

\end{document}